\documentclass[12pt]{article}%
\usepackage{amssymb}
\usepackage{float}
\usepackage{epsfig, graphicx}
\usepackage{amsmath}
\usepackage{amsfonts}
\usepackage{graphicx}%
\newtheorem{theorem}{Theorem}
\newtheorem{lem}{Lemma}

\newtheorem{prop}{Proposition}

\begin{document}

\author{David Holcman \thanks{Department of Mathematics, Weizmann Institute of
Science, Rehovot 76100 Israel. Visiting ad. Department of Mathematics, University of
California at Berkeley, CA 94720-3860 and Keck-Center, Department of Physiology, UCSF
513 Parnassus Ave, San Francisco CA 94143-0444, USA. I would like to thank Pr. G. Tomassini for his financial support: part of this work was done in summer 2001 at SNS, Pisa.} \and Ivan Kupka\thanks{ Universit\'e Paris VI, department of Mathematic, 175 rue du
Chevaleret 75013 Paris, France.}}
\title{Perturbation Methods and First Order Partial Differential Equations}
\date{November 26/2002}
\maketitle

\begin{abstract}
In this paper, we give explicit estimates that insure the existence of
solutions for\ first order partial differential operators on compact
manifolds, using a viscosity method. In the linear case, an explicit integral
formula can be found, using the characteristics curves. The solution is given
explicitly on the critical points and the limit cycles of the vector field of
the first order term of the operator. In the nonlinear case, a generalization
of the Weitzenb\"{o}ck formula provides pointwise estimates that insure the
existence of a solution, but the uniqueness question is left open.
Nevertheless we prove that uniqueness is stable under a C$^{1}$ perturbation.
Finally, we give some examples where the solution fails to exist globally,
justifying the need to impose conditions that warrant global existence. The
last result reveals that the zero order term in the first order operator is
necessary to obtain generically bounded solutions.

\end{abstract}
\tableofcontents

\section{\bigskip Introduction}

\ 

In this paper, we consider a compact orientable\ Riemannian manifold (V,g), a
smooth vector field b on V or a parametrized vector field b:$\mathbb{R}$xV---%
$>$%
TV, the tangent bundle of V, and a smooth positive function c:V---%
$>$%
$\mathbb{R}$ or a parametrized positive smooth function c:$\mathbb{R}$xV---%
$>$%
$\mathbb{R}$. $\Delta_{g}$ is the negative Laplacian:
\[
\Delta_{g}=-div_{g}grad_{g}%
\]
where div$_{g}$ denotes the divergence operator with respect to the volume
form associated to the metric g and grad$_{g}$ the gradient with respect to g.

We shall study the limit of the solutions $u_{\epsilon}$ as $\varepsilon$
tends to 0 through positive values of the linear equations:
\begin{equation}
\epsilon\Delta_{g}u_{\epsilon}+<b,gradu_{\varepsilon}>_{g}+cu_{\epsilon
}=\text{f on V} \label{el}%
\end{equation}
where $f$ is a given continuous function on the manifold V or of the non
linear equations:
\begin{equation}
\epsilon\Delta_{g}u_{\epsilon}(x)+<b(u_{\varepsilon}(x),x),gradu_{\epsilon
}(x)>_{g}+c(%
\operatorname{u}%
_{\varepsilon}(x),x)u_{\varepsilon}(x)=0\text{, x}\in\text{V} \label{fdtf}%
\end{equation}

\bigskip Heuristically, the limits if they exists, ''should'' be solutions of
the first order partial differential equation:
\begin{equation}
<b,gradu>_{g}+cu=\text{f on V} \label{eql}%
\end{equation}
in the linear case and:
\begin{equation}
<b(u(x),x),gradu(x)>_{g}+c(%
\operatorname{u}%
(x),x)u(x)=0,x\in V \label{eq}%
\end{equation}
$\bigskip$in the nonlinear case. The linear case is obviously a particular
case of the non linear one.

Historically, Cauchy devised a powerful method to find local solutions of the
Cauchy or initial value problem for first order partial differential equations
(linear or non linear) using the characteristic curves which were the
solutions of an ordinary differential system in the first jet bundle, called
the characteristic system (see \cite{CH}). The initial data were given on a
hypersurface in the bundle of the first jets of functions on the space of the
partial differential equation and then propagated along the characteristic
curves from the hypersurface. In the case of equation (\ref{eq}) , using the
natural coordinate system (x$^{1}$,...,x$^{m}$,u,p$_{1}$,...,p$_{m}$), on the
jet space, the characteristic system for equation(\ref{eq}) can be expressed
as follows:
\begin{align}
\frac{dx^{n}}{dt}  &  =b^{n}(u,x)\label{char}\\
\frac{du}{dt}  &  =\sum_{k=1}^{m}p_{k}b^{k}(u,x)\nonumber\\
\frac{dp_{n}}{dt}  &  =-\sum_{k=1}^{m}\left(  \frac{\partial b^{k}}{\partial
x^{n}}(u,x)+p_{n}\frac{\partial b^{k}}{\partial u}(u,x)\right)  p_{k}%
-p_{n}(u\frac{\partial c}{\partial u}(u,x)+c(u,x))-\frac{\partial c}{\partial
x^{n}}(u,x)\nonumber
\end{align}
where $b^{k}(u,x),$1$\leq k\leq m,$ are the components of the field b.

The problem of solving (\ref{eq}) is not an initial value problem but a
Dirichlet problem. It is not well behaved because the differential operator
appearing in the left hand-side of equation(\ref{eq}) is not elliptic.
Nevertheless the Cauchy characteristics are still very useful for the study of
equation(\ref{eq}). Indeed in the linear case \ one can give an explicit
formula for the solution u of equation(\ref{eql}) :
\begin{equation}
\forall P \in V, u(P)=\int_{-\infty}^{0}f(x_{P}(s))e^{-\int_{s}^{0}%
c(x_{P}(\tau))d\tau} \label{form1}%
\end{equation}
see formula(\ref{form}) in theorem(\ref{thepdf}) below. Actually this formula
can be used to prove the existence and properties of classical (i.e.at least
Lipschitz continuous) solutions to equation(\ref{eql}) due to the fact that
the equation(\ref{char}) can be solved without the knowledge of the solution
u. Actually, provided that $\underset{V}{\min}c$%
$>$%
0, the formula (\ref{form1}) defines a function u on V but this function is,
in general, not even continuous. In the non linear case, the
formula(\ref{form1}) is replaced by the integral equation in u:
\begin{align*}
u(x)  &  =\int_{-\infty}^{0}K(t,x)dt,\text{ \ \ x}\in V,\text{ \ where}\\
\text{K(t,x)}  &  =\text{f(}\varphi^{u}(t,x))\exp[-\int_{t}^{0}c(u(\varphi
^{u}(s,x)),\varphi^{u}(s,x))ds]
\end{align*}
see equation(\ref{integ}) below. This integral equation can be solved using the
Picard fixed point theorem and give us a Lipschitz continuous solutions under
assumptions which are stronger than the ones needed to prove the existence
using the viscosity method developed below.

The standard viscosity method is the approach using an elliptic partial
differential equation to find solutions for a first order partial differential
equation. Other methods such as the one mentioned in the last paragraph, exist
but the advantage of the viscosity method is that, explicit geometrical
conditions and explicit constants in inequalities are derived. Under these
explicit conditions, the existence and local uniqueness of solutions are
established. We prove that in certain types of first order linear partial
differential equation, the solution is unique and a representation of this
solution can be found using the trajectories of a vector field.

First order partial differential equations similar to equation \ref{eq} and \ref{eql}
 were considered in relation to the KAM theory by Kuksin in \cite{kuksin}. Here
V is a m-dimensional torus endowed with the flat metric b is a
translation-invariant purely imaginary vector field. c must satisfies the
condition (dx is the Haar measure)
\begin{equation}
\left\vert \int_{V}c(x)dx\right\vert >>\max\{\text{osc(c),1}\}\label{kuks}.%
\end{equation}
Under some incommensurability assumptions on the coefficients of the field in
the standard basis of the lie algebra of the torus and the condition
\ref{kuks}, Kuksin proves that equation \ref{eql} has a unique
analytic solution for any analytic f. Moreover sup-norm estimates of the
solution are given in terms of the sup-norm of f. The method of proof is based
on the theory of similar equations on the circle S$^{1}$. Because we are on a
torus Fourier series can be used to represent the functions and this is how
the estimates are derived. 

Forni in \cite{forni} studies equation \ref{eql} with c=0 on compact
surfaces endowed with a Riemannian metric. He assumes that the vector field b
preserves the volume measure associated to the metric and that its
singularities are nondegenerate saddle points only. When f is sufficiently
regular and its mean value vanishes, there exists a unique solution( up to a
constant of course).

In the following paper, there are no restrictions on the underlying manifold V
(except compactness), b is a general Morse-Smale vector field. Hence in
particular it is not volume preserving. Also in our case c cannot be taken to
be 0. It has to satisfy condition $\underset{V}{\min}$ c%
$>$
b$_{0}$(defined in the notations below) which is more stringent than
\ref{kuks}. This condition means that c must be larger than the variation of
b. In the case considered by Kuksin, this variation is 0 because b is
self-parallel.\qquad

Some of the results exposed here were announced in \cite{HK1}. Let us finally
mention that when the function f in the right hand side of equation
(\ref{fdtf}) is proportional to $u$, we get an eigenvalue problem. The
analysis of this problem  has been done in the general case in \cite{HK2,HK3}.

The paper is organized as follow. In the first section of the paper we study
the limits as $\varepsilon$ goes to 0 of the solutions of equation (\ref{el})
to prove the existence and uniqueness of solutions for the linear equation
(\ref{eql}). Then we prove the uniqueness of the solutions. In the second
section \ we take up the existence and uniqueness of solutions of the
nonlinear elliptic equation (\ref{fdtf}). The estimates needed to achieve this
are provided by the Weitzenb\"{o}ck formula and its generalization to
covariant tensors of valence two. Then as in the first section we examine the
limits of solutions of equation (\ref{fdtf}) as $\varepsilon$ goes to 0 and
prove the existence of solutions for the nonlinear first order equation
(\ref{eq}). In the third and last section, we give various examples in the
nonlinear case where a solution can exists in an open set but not globally.
When the field is ergodic, we give a simple result about the behavior of the
solution on the manifold. Finally in order to understand how relevant is the
zero order term, we consider the case where it is zero. In that case, the
viscosity method does not converge to any bounded solutions. Generically, a
first order linear partial differential equation with no zero order term
cannot have bounded solutions.

\subsection{Notations}

\bigskip Throughout this paper we shall use the following notations: \ 

\ 1){\large Metric and tensors:} \ %

\begin{align*}
d_{g} &  =\text{distance function defined on V by the metric }\\
||\ast||, &  <\ast,\ast>_{g}=\text{norm and scalar product associated to g}\\
vol_{g} &  =\text{volume measure associated to g}\\
grad &  =\text{gradient operator associated to g}\\
div_{g} &  =\text{divergence operator associated to g}\\
\nabla &  =\text{Levi-Civita covariant derivative associated to g}\\
R &  =\text{curvature tensor of }\\
\theta(b) &  :=\text{ Lie derivative operator associated to the vector field
b}\\
\theta(b)u &  =du(b)
\end{align*}
For any coordinate system x$^{1}$,....,x$^{m}:\mathcal{O}$---%
$>$%
$\mathbb{R}$ on V :
\begin{align*}
g &  =g_{ij}dx^{i}\otimes dx^{j}\\
g^{ij} &  =\text{inverse of the matrix g}_{ij}\\
vol_{g} &  =\sqrt[2]{\text{det(g}_{ij})}dx^{1}...dx^{m}\\
gradu^{i} &  =g^{ij}\frac{\partial u}{\partial x^{j}}\\
\nabla_{i} &  =\nabla_{\frac{\partial}{\partial x^{i}}}\\
div_{g}X &  =\nabla_{i}X^{i}\\
R &  =R_{ijk\cdot}^{\cdot\cdot\cdot l}dx^{i}\otimes dx^{j}\otimes
dx^{k}\otimes\frac{\partial}{\partial x^{l}}\\
\nabla_{i}\nabla_{j}-\nabla_{j}\nabla_{i} &  =R_{ijk\cdot}^{\cdot\cdot\cdot
l}dx^{k}\otimes\frac{\partial}{\partial x^{l}}%
\end{align*}
If $\mathcal{T}$ is a p-covariant,q-contravariant vector field:%
\[
\mathcal{T=T}_{i_{1},..i_{p}}^{\bullet,...\bullet j_{1},..,j_{q}}dx^{i_{1}%
}\otimes...\otimes dx^{i_{p+1}}\otimes e_{j_{1}}\otimes...\otimes e_{j_{q}}%
\]%
\[
\nabla\mathcal{T}=\nabla_{i_{1}}\mathcal{T}_{i_{2},..i_{p+1}}^{\bullet
,...\bullet j_{1},..,j_{q}}dx^{i_{1}}\otimes...\otimes dx^{i_{p+1}}\otimes
e_{j_{1}}\otimes...\otimes e_{j_{q}}%
\]
where (e$_{1}$,...,e$_{m}$) is the frame field associated to the coordinates
x$^{1}$,....,x$^{m},$ on $\mathcal{O}$. \bigskip

2){\large Norms:} g induces a scalar product function and a norm \ function on
any tensor bundle on V. Let t, $\tau$ be two tensors of the same type, at the
same point of V.
\begin{align*}
<t,\tau>_{g}  &  =\text{scalar product of t and }\tau\\
||t||_{g}\text{{}}  &  =\text{norm of t}%
\end{align*}
To any tensor field $\mathcal{T}$ \ on V\ is associated the function:

x$\in V$---%
$>$%
$\vert$%
$\vert$%
$\mathcal{T}$\
$\vert$%
$\vert$%
$_{g}\in\lbrack0,+\infty\lbrack.$%
\begin{align*}
||\mathcal{T}||_{\infty}  &  =\underset{x\in V}{\max}||\mathcal{T}\text{ (x)%
$\vert$%
$\vert$%
}_{g}\\
||\mathcal{T}||_{L^{\infty}}  &  =\underset{x\in V}{\text{ess-sup}%
}||\mathcal{T}\text{ (x)%
$\vert$%
$\vert$%
}_{g}\\
||\mathcal{T}||_{L^{2}}  &  =\left(  \int_{V}||\mathcal{T}\text{ (x)%
$\vert$%
$\vert$%
}_{g}^{2}vol_{g}(dx)\right)  ^{\frac{1}{2}}\\
||\mathcal{T}||_{C^{1}}  &  =||\mathcal{T}||_{\infty}+||\nabla\mathcal{T}%
\text{
$\vert$%
$\vert$%
}_{\infty}\\
||\mathcal{T}||_{W^{1,\infty}}  &  =||\mathcal{T}||_{L^{\infty}}%
+||\nabla\mathcal{T}\text{
$\vert$%
$\vert$%
}_{L^{\infty}}\\
||\mathcal{T}||_{H^{1}}  &  =\sqrt[2]{||\mathcal{T}||_{L^{2}}^{2}%
+||\nabla\mathcal{T}||_{L^{2}}^{2}}%
\end{align*}
\bigskip

3$)${\large Constants:}
\begin{align*}
b_{0}  &  =\sup\{<\nabla_{X}b,X>_{g}|X\in TV,||X||_{g}=1\}\\
\beta &  =||\frac{\partial b}{\partial\lambda}||_{\infty}=\underset
{\mathbb{R\times V}}{\max}||\frac{\partial b}{\partial\lambda}||_{g}\\
c_{0}  &  =\inf_{\lambda\in\mathbb{R}\text{,}x\in V}c(\lambda,x)>0\\
f_{0}  &  =||df||_{\infty}+\frac{||f||_{\infty}||dc||_{\infty}}{c_{0}}\\
\gamma &  =||\frac{\partial c}{\partial\lambda}||_{\infty}.\frac
{||f||_{\infty}}{c_{0}}\\
r_{0}  &  =\max\{|Ricc(\omega,\omega)|\text{ ,}|\text{ }\omega\in\text{T*V,%
$\vert$%
$\vert$%
}\omega||_{g}=1\}
\end{align*}
\bigskip

4){\large Lipschitz properties}

A function f:V---%
$>$%
$\mathbb{R}$, will be called Lipschitz continuous if :
\[
\underset{x,y\in V}{\sup}\frac{|f(x)-f(y)|}{d_{g}(x,y)}<+\infty
\]
The sup is called the Lipschitz constant of f and will be denoted by Lip(f).
Any upper \ bound of the Lipschitz constant is called a Lipschitz bound of f.
Finally we denote%

\begin{align*}
C^{0,1}(V)  &  =\text{Banach space of all lipschitz continuous functions on V
with the norm}\\
||f||_{C^{0,1}}  &  =\underset{V}{\max}|f|+Lip(f)
\end{align*}

\section{\bigskip The linear case:main theorem}

In this section, c:V---%
$>$%
$\mathbb{R}$, will denote a smooth positive function and f:V---%
$>$%
$\mathbb{R}$, a Lipschitz continuous one. Let us recall that by the theory of
second order elliptic equations(\cite{BJS}), the following equation:
\begin{equation}
\epsilon\Delta_{g}u_{\epsilon}+<b,gradu_{\varepsilon}>_{g}+cu_{\epsilon
}=\text{f on V} \label{equa}%
\end{equation}
has a unique solution u$_{\varepsilon}$ which is of any class C$^{2,\alpha} $,
for $\alpha\in\lbrack0,1[. $ If f is of class C$^{k}$, k$\in\mathbb{N\cup
\infty}$, u$_{\varepsilon}$ is of class C$^{k+2}$. We shall study the limit
when $\epsilon$ converges to zero of the\ solution $u_{\epsilon}$of the
equation(\ref{equa}).

\begin{theorem}
\label{theex}

\begin{enumerate}
\item On a compact manifold, consider a smooth vector field b, a smooth
positive function c satisfying $\underset{V}{\min}c=c_{0}>b_{0}$ and a
Lipschitz continuous function f. Under these assumptions, there exists a
unique Lipschitz-continuous function u:V---%
$>$%
$\mathbb{R}$, solution of the first order partial differential equation :
\begin{equation}
<b,gradu>_{g}+cu=\text{f on V} \label{edpline}%
\end{equation}
in the weak sense.

\item For any P$\in V,$ if x$_{P}$:$\mathbb{R}\rightarrow V$, denotes the
trajectory of b such that x$_{P}$(0)=P,we have the formula:
\begin{equation}
u(P)=\int_{-\infty}^{0}f(x_{P}(s))e^{-\int_{t}^{0}c(x_{P}(s))ds}dt
\label{form}%
\end{equation}

\item The solution u$_{\varepsilon}$ of \ the equation $\epsilon\Delta
_{g}u_{\epsilon}+<b,\nabla u_{\epsilon}>+cu_{\epsilon}=f$ \ tends to u in the
sup norm topology\ as $\varepsilon$ tends to zero.

\item If f is of class C$^{1}$ this solution is also C$^{1}$.

\item At a singular point P of b, u(P)=$\frac{f(P)}{c(P)}$.

\item If p:$\mathbb{R}\rightarrow V_{n}$ is a periodic trajectory of b with
minimal period T, for all t$\in\mathbb{R}$%
\begin{equation}
u(p(t))=\frac{C(T)}{1-C(T)}\int_{0}^{t}f(p(s))\frac{C(t)}{C(s)}ds+\frac
{1}{1-C(T)}\int_{t}^{T}f(p(s))\frac{C(t)}{C(s)}ds,\text{ for all t},
\label{limitC}%
\end{equation}
\ \ \ \ where C(t)=$\exp[-\int_{0}^{t}c(p(s))ds].$
\end{enumerate}
\end{theorem}

\textbf{\noindent Proof :} First we will prove that any sub-sequences of
$u_{\epsilon}$ contains a converging sub-sequence whose limit u in the weak
L$_{2}$ topology, is a Lipschitz continuous function satisfying equation
(\ref{edpline}). In order to show this we need to establish some a priory
estimates which follow easily from the maximum principle. The following is
easy:
\begin{equation}
\frac{\underset{V}{\min f}}{\underset{V}{\max c}}\leq u_{\epsilon}\leq
\frac{\underset{V}{\max f}}{\underset{V}{\min c}} \label{primax}%
\end{equation}
There exists a sub-sequence of $u_{\epsilon}$ which converges weakly to a
function $u$ and for all $\phi\in H_{1}$,
\[
\epsilon\int_{V}u_{\epsilon}\Delta_{g}\phi-\int_{V}div(\phi b)u_{\epsilon
}+\int_{V}\phi cu_{\epsilon}=\int_{V}\phi f\text{ }%
\]
$u$ satisfies for all $\phi\in H_{1}$,
\[
-\int_{V}div(\phi b)u+\int_{V}\phi cu=\int_{V}\phi f
\]
In these equations the integrations are with respect to the measure defined by
the metric g on V.

Now we will prove that
$\vert$%
$|du_{\epsilon}||_{\infty}$ is also bounded. Let us note here that using a
stronger lower bound on c, one could prove that $||\nabla du_{\epsilon}||_{g}$
is also bounded in $L_{\infty}$ which would imply differentiability of the
limit. To estimate
$\vert$%
${|du_{\epsilon}||}_{\infty}^{2}$, we shall consider the maximum of the
function
$\vert$%
${|du_{\epsilon}||}^{2}$, following a standard method (see
\cite{Moser,Kamin,Sacker} ).

Starting with the Weitzenb\"{o}ck formula:
\[
\frac{1}{2}\Delta_{g}||{du_{\epsilon}||}_{g}^{2}+||{\nabla du_{\epsilon}%
||}_{g}^{2}+Ric(du_{\epsilon},du_{\epsilon})=<d\Delta_{g}u_{\epsilon
},du_{\epsilon}>_{g}%
\]
Assuming that f is C$^{1}$, an easy computation gives :%

\begin{align*}
&  \epsilon{\ }\frac{1}{2}{\Delta}_{g}{(||du}_{\epsilon}|{|}_{g}^{2}%
)+\epsilon||\nabla du_{\epsilon}||_{g}^{2}+\epsilon Ricci(du_{\epsilon
},du_{\epsilon})+c||d{u_{\epsilon}||}_{g}^{2}+<\nabla_{gradu_{\epsilon}%
}b,gradu_{\epsilon}>_{g}\\
&  =-\frac{1}{2}\theta(b)||du_{\epsilon}||_{g}^{2}+<df,du_{\epsilon}%
>_{g}-u_{\epsilon}<dc,du_{\epsilon}>_{g}.
\end{align*}
Because the Ricci curvature is bounded from below by $r_{0}$ and $u_{\epsilon
}$ is bounded from above , we obtain that at a maximum point P of the
function
$\vert$%
${|du_{\epsilon}||}_{g}^{2}$,
\[
(c_{0}-\epsilon r_{0}-b_{0})||{du_{\epsilon}(P)||}_{g}^{2}\leq\mid
<df(P),du_{\epsilon}(P)>_{g}|+|u_{\epsilon}(P)<dc(P),du_{\epsilon}(P)>_{g}|
\]
Note that at a maximum point P of the function ${\ ||du_{\epsilon}||}_{g}^{2}%
$, ${\ \Delta}_{g}{(||du_{\epsilon}||}_{g}^{2})(P)\geq0$ and ($\theta
(b)||du||_{g}^{2})(P)=0. $ Hence we get using the relations (\ref{primax}):
\begin{equation}
{\ ||du_{\epsilon}||}_{\infty}=||{du_{\epsilon}(P)||}_{g}\leq\frac
{||df||_{\infty}+\frac{||f||_{\infty}||dc||_{\infty}}{c_{0}}}{c_{0}-\epsilon
r_{0}-b_{0}} \label{lipin}%
\end{equation}
Now we drop the auxiliary assumption that f$\in C^{1}$and prove that for any
Lipschitz continuous f:
\[
{||du_{\epsilon}||}_{\infty}\leq\frac{||df||_{L^{\infty}}+\frac{||f||_{\infty
}||dc||_{\infty}}{c_{0}}}{c_{0}-\epsilon r_{0}-b_{0}}%
\]
We use the following lemma (probably well known) proved in the appendix:

\begin{lem}
\label{extlip}\bigskip Let k:V---%
$>$%
$\mathbb{R}$, be a Lipschitz continuous function with M as Lipschitz bound.
Then for any neighborhood $\mathfrak{U}$ of k in C$^{0}$(V), any
$\varepsilon>0$, there exists a C$^{\infty}$ function h:V---%
$>$%
$\mathbb{R}$, contained in $\mathfrak{U}$ and admitting M+$\varepsilon$,as
Lipschitz bound.
\end{lem}

\bigskip Hence with the lemma, we can find a sequence \{f$_{n}|n\in
\mathbb{N\}}$ of smooth functions on V, converging to f in C$^{0}$(V), f$_{n}$
allowing $||df||_{L^{\infty}}+\frac{1}{n}$ as Lipschitz bound for each n. Let
u$_{\epsilon,n}$ be the solution of equation (\ref{equa}) with second member
f$_{n}$. It follows from classical elliptic estimates that the sequence
\{u$_{\varepsilon,n}|n\in\mathbb{N\}}$ converges to u$_{\varepsilon}$ in the
C$^{2}$ topology. The inequality (\ref{lipin}) implies that for each
n$\in\mathbb{N}$:
\[
{\ ||du_{\epsilon,n}||}_{\infty}\leq\frac{||df_{n}||_{\infty}+\frac
{||f_{n}||_{\infty}||dc||_{\infty}}{c_{0}}}{c_{0}-\epsilon r_{0}-b_{0}}%
\]
for every $\varepsilon\in\lbrack0,\frac{c_{0}-b_{0}}{2}]$. From this it
follows that u$_{\varepsilon}$ is Lipschitz continuous and that:
\[
{\ ||du}_{\varepsilon}{||}_{\infty}\leq\frac{||df||_{L^{\infty}}%
+\frac{||f||_{\infty}||dc||_{\infty}}{c_{0}}}{c_{0}-\epsilon r_{0}-b_{0}}%
\]
Because $||u_{\epsilon}||_{g}$ ,$||du_{\epsilon}||_{g}$ are uniformly bounded
in $\epsilon$ for $\varepsilon\in\lbrack0,\frac{c_{0}-b_{0}}{2}]$, any
sequence contains a sequence converging uniformly and even in any H\"{o}lder
space $C^{0,\alpha}$, for $\alpha<1$, to a solution u of the equation
\[
<b(x),gradu(x)>_{g}+c(x)u(x)=f(x)\text{ almost everywhere on V}%
\]
Also
$\vert$%
$\vert$%
du%
$\vert$%
$\vert$%
$_{L^{\infty}}\ $is finite and:
\[
{||du||}_{L^{\infty}}\leq\frac{||df||_{L^{\infty}}+\frac{||f||_{\infty
}||dc||_{\infty}}{c_{0}}}{c_{0}-b_{0}}.
\]
\textbf{\noindent}

\subsection{proof of existence and uniqueness}

Now we prove that there is only one function u:$V\rightarrow\mathbb{R}$, which
is a weak solution of equation(\ref{edpline}) and is Lipschitz continuous. We
have just shown that such u's exist. Taking any one of them we will give an
integral representation of the function u along the trajectories of b which
will be used to compute the value of the function $u$ at the stationary points
and along the periodic trajectories of b. Since $u$ is Lipschitz continuous,
it is absolutely continuous along any $C^{1}$ curve and almost everywhere
differentiable. If x:$\mathbb{R}\rightarrow V$ is any trajectory of b, the
function: t$\in\mathbb{R}\rightarrow u(x(t))$ is absolutely continuous and
satisfies the equation:%
\begin{equation}
\frac{du(x(t))}{dt}+c(x(t))u(x(t))=f((x(t))\ \ \text{almost everywhere}
\label{simpl}%
\end{equation}
Hence the function:t$\in\mathbb{R}\rightarrow u(x(t))$ is a genuine C$^{1}$
solution of equation(\ref{simpl}) and we have for all t, t$_{1}$,
\[
u(x(t))=u(x(t_{1}))e^{-\int_{t_{1}}^{t}c(x(s))ds}+\int_{t_{1}}^{t}%
f(x(s))e^{-\int_{s}^{t}c(x(\tau))dt}ds
\]
\ \ \ \ \ Because u is bounded and $\underset{V}{\min}c\geqq c_{0}>0,$ we get,
if we let \ t$_{1}$ \ go to -$\infty$ :
\[
u(x(t))=\int_{-\infty}^{t}f(x(s))e^{-\int_{s}^{t}c(x(\tau))d\tau}%
\]
For any point P in V, if x$_{P}$:$\mathbb{R}$---%
$>$%
V denotes the trajectory of b passing through P at time 0:
\[
u(P)=\int_{-\infty}^{0}f(x_{P}(s))e^{-\int_{s}^{0}c(x_{P}(\tau))d\tau}%
\]
\bigskip This formula proves the uniqueness of u. It also proves the assertion 3)

\subsection{ proof of regularity}

\bigskip\ \bigskip We need an elementary lemma:

\begin{lem}
\label{lemmff}

\begin{enumerate}
\item If $\rho$ is the modulus of an eigenvalue of the linearization of the
flow at time one of b at a singular point of b, then $\mid log\rho\mid\leq
b_{0}$.

\item If $\rho$ is the modulus of an eigenvalue of the monodromy of a periodic
orbit of b of minimal period T, then $\mid log \rho| \leq b_{0}T $.

\item If $\{\phi_{t}\mid t\in\mathbb{R}$\} denotes the flow of b and
T$\phi_{t}$ its tangent mapping, then $\mid T\phi_{t}\mid\leq e^{b_{0}|t|}$
for all real t.
\end{enumerate}
\end{lem}

\bigskip\noindent\textbf{Proof of the lemma:} The first statement is easy and
left to the reader. Let us prove 3 first. Denote by v$_{0}$ any tangent vector
to V. Let $v:\mathbb{R}\rightarrow TV$, be the vector field :
\[
v(t)=T\phi_{t}\text{(v}_{0}\text{)}%
\]
We have:
\begin{align*}
\nabla_{t}v  &  =\nabla_{v}b\\
\frac{d||v(t)||_{g}^{2}}{dt}  &  =2<\nabla_{v(t)}b,v(t)>_{g}\\
\frac{d||v(t)||_{g}^{2}}{dt}  &  \leq2b_{0}||v(t)||_{g}^{2}%
\end{align*}
From this it follows that if t$\geq0$:
\[
||v(t)||_{g}\leq||v(0)||_{g}e^{b_{0}t}%
\]
Reversing the time, we get for $t<0$:
\[
||v(t)||_{g}\leq||v(0)||_{g}e^{b_{0}|t|}%
\]
This last inequality implies 3).

\noindent Let us prove 2). Let p:$\mathbb{R}\rightarrow V$ be\ a periodic
trajectory of b of (minimal) period T and let $\lambda$ be an eigenvalue of
the monodromy of p. There exists a tangent vector v$_{0}\in$T$_{p(0)}%
$V$\otimes\mathbb{C}$, v$_{0}\neq0$, such that T$\phi_{T}$(v$_{0}$)=$\lambda
$v$_{0}$ and T$\phi_{-T}$(v$_{0}$)=$\frac{1}{\lambda}$v$_{0}$. By the above:
\begin{align*}
|\lambda|  &  \leq e^{b_{0}T}\\
\frac{1}{|\lambda|}  &  \leq e^{b_{0}T}%
\end{align*}
This two inequalities imply assertion 2).\bigskip

\bigskip Now we resume the proof of assertion 4). Let \{$\phi_{t}%
|t\in\mathbb{R\}}$ denote the flow of b. $\varphi(t,x)=\varphi_{t}(x)$.

\noindent For any x$\in$V:
\begin{equation}
u(x)=\int_{-\infty}^{0}K(t,x)dt \label{ex}%
\end{equation}
where K:$\mathbb{R}\times V\rightarrow\mathbb{R}$, is the function:
\[
K(t,x)=f(\phi(t,x))e^{-\int_{t}^{0}c(\phi(s,x))ds}%
\]
Let K$_{t}(x)=$K(t,x).K is a C$^{\infty}$ function. Its differential in x is:
\[
dK_{t}(x)=e^{-\int_{t}^{0}c(\phi(s,x))ds}[df(\phi(t,x))T_{x}\phi_{t}-\int
_{t}^{0}dc(\phi(s,x))T_{x}\phi_{s}ds]
\]
$T_{x}\phi_{t}$ is the tangent mapping of the diffeomorphism $\varphi_{t}%
$:V---%
$>$%
V, at x. The second integral is that of the curve: s$\in\mathbb{R}\rightarrow
dc$($\phi$(s,x))T$_{x}\phi_{s}\in$T$_{x}^{\ast}$V, the cotangent space of V at
x. Using the statement 3) of the Lemma(\ref{lemmff}), for t$\leq0$:
\[
||dK_{t}(x)||\leq e^{(c_{0}-b_{0})t}[||f||_{1}+|t|||c||_{1}]
\]
This estimate, uniform for all x $\in$X and all t$\leq0$, shows that u is
continuously differentiable on V because of the assumption \ c$_{0}>$b$_{0}$
and that:
\[
du(x)=\int_{-\infty}^{0}dK(t,x)dt
\]

\subsection{explicit formulas on the recurrent set}

\bigskip\noindent Now we can specialize the formula(\ref{ex}) to different
kinds of trajectories. Let P be a singular point of b. Then the curve
x:$\mathbb{R}\rightarrow V,x(t)=P$ for all t is a trajectory and
\[
u(P)=u(x(0))=\int_{-\infty}^{0}f(x(s))e^{-\int_{s}^{0}c(x(\tau))d\tau}%
=\int_{-\infty}^{0}f(P)e^{-\int_{s}^{0}c(P)d\tau}=\frac{f(P)}{c(P)}.
\]
\bigskip\noindent Now assume that p:$\mathbb{R}\rightarrow V$, is a periodic
trajectory of (minimal) period T. Applying the general formula to p, we get
\[
u(p(t))=\int_{-\infty}^{t}f(p(s))e^{-\int_{s}^{t}c(p(\tau))d\tau}%
\]
We choose t$\in$[0,T]. Then:
\begin{align*}
u(p(t))  &  =C(t)[\int_{-\infty}^{0}\frac{f(p(s))}{C(s)}ds+\int_{0}^{t}%
\frac{f(p(s))}{C(s)}ds]\\
\int_{-\infty}^{0}\frac{f(p(s))}{C(s)}ds  &  =\sum_{n=0}^{+\infty}%
\int_{-(n+1)T}^{-nT}\frac{f(p(s))}{C(s)}ds\\
\int_{-(n+1)T}^{-nT}\frac{f(p(s))}{C(s)}ds  &  =e^{-(n+1)C(T)}\int_{0}%
^{T}\frac{f(p(s))}{C(s)}ds\\
\int_{-\infty}^{0}\frac{f(p(s))}{C(s)}ds  &  =\frac{e^{-C(T)}}{1-e^{-C(T)}%
}\int_{0}^{T}\frac{f(p(s))}{C(s)}ds\\
u(p(t))  &  =C(t)[\frac{e^{-C(T)}}{1-e^{-C(T)}}\int_{0}^{T}\frac
{f(p(s))}{C(s)}ds+\int_{0}^{t}\frac{f(p(s))}{C(s)}ds]\\
u(p(t))  &  =C(t)[\frac{e^{-C(T)}}{1-e^{-C(T)}}\int_{t}^{T}\frac
{f(p(s))}{C(s)}ds+\frac{1}{1-e^{-C(T)}}\int_{0}^{t}\frac{f(p(s))}{C(s)}ds]
\end{align*}
\quad\hbox{\hskip 4pt\vrule width 5pt height 6pt depth 1.5pt} \quad

\noindent\textbf{Remark 1 :} Let us define the following real number:
\[
B_{0}=\inf_{g}b_{0}(g)=\inf_{g\in Riem}\max\{|<\nabla_{X}b,X>_{g}|X\in{TV,\mid
X\mid=1\}},
\]
where Riem denotes the Riemannian structure on V. Lemma \ref{lemmff} shows
that $B_{0}$ is strictly positive (bigger than the minimum of the log$\rho$
and $\frac{ \log\rho}{T}$ ). For each function c such that $\underset{V}{\inf
}c$%
$>$%
B$_{0}$, the formula(\ref{ex}) defines a function $u_{c}$ which is a regular
solution of equation(\ref{edpline}): indeed one chooses a Riemannian metric g
on V such that $\underset{V}{\inf}c>b(g)$ and applies what has just been
proved. When $\underset{V}{\inf}c$= $B_{0}$, the formula (\ref{ex}) still
defines a function u$_{c}$, but it is not even sure that it is continuous,
even less that it is a solution of equation (\ref{edpline}) in some sense.
Hence the following question: what happens in this situation? What type of
solutions, if any does the equation(\ref{edpline}) has ?

\noindent\textbf{Remark 2 :} Can one find an explicit expression for the limit
of the solution u, on a general recurrent set? what about the case, when the
recurent sets have an invariant measure? this will extend the formula obtained
on the limit cycles.

\section{ The nonlinear case}

In this section, we consider a parametrized vector field :$\lambda
\in\mathbb{R}$---%
$>$%
b$_{\lambda}\in$VF(V), space of all vector fields on V. b:$\mathbb{R}$xV---%
$>$%
TV will denote the mapping: ($\lambda,x)$---%
$>$%
b$(\lambda,x)=b_{\lambda}(x).$c: $\mathbb{R}$xV---%
$>$%
$\mathbb{R}$, will denote a smooth function and c$_{\lambda}$:V---%
$>$%
$\mathbb{R}$ will be the function c$_{\lambda}(x)=c(\lambda,x).$

We shall study the following equation on V:\ \ \
\begin{equation}
<b_{u},gradu>+c_{u}=f\text{ on }V \label{eedp}%
\end{equation}
where u:V---%
$>$%
$\mathbb{R}$, is the unknown function, f:V---%
$>$%
$\mathbb{R}$, is a given smooth function on V, b$_{u}$ is the vector field on
V,b$_{u}$(x)=b(u(x),x), c$_{u}$ the function\ c$_{u}$(x)=c(u(x),x).

We will impose some conditions on $f,c$ and $b$ in order to insure the
existence of regular solutions. Let us note here that there are examples with
discontinuous solutions. This has been discussed in \cite{JKM}. To prove the
existence of solutions for equation (\ref{eedp}), we use a standard elliptic
regularization techniques and proceed by successive approximations.

The following two assumptions will be in force in the following sections:


\begin{enumerate}
\item $c_{0}-b_{0}-\gamma>0.$ This implies that c$_{0}$%
$>$%
0.

\item ($c_{0}-b_{0}-\gamma)^{2}-4f_{0}\beta>0$
\end{enumerate}

These assumptions say that the minimum of the function $c$ must be large
enough compared to the maximum dilation of the field b and are called
conditions of hyperbolicity.


\subsection{Existence of the solution for the elliptic PDE}

To prove the first part of the theorem, about the existence of a solution to
the first order partial differential equation, we will built a sequence of
solutions of some nonlinear elliptic partial differential equations with a
small viscosity coefficient and later on, we will let this small parameter
converges to zero.

Now the assumptions (1-2) imply that there exists two strictly positive roots
of the second order equation: $\beta X^{2}-X(c_{0}-b_{0}-\gamma)+f_{0}=0$. The
same property will be true for the equation: $\beta X^{2}-X(c_{0}%
-b_{0}-\varepsilon r_{0}-\gamma)+f_{0}=0$ provided that 0$\leq$ $\varepsilon
\leq\overline{\varepsilon}$, where
\begin{equation}
\overline{\varepsilon}\text{ =}\frac{c_{0}-b_{0}-\gamma-2\sqrt{f_{0}\beta}%
}{r_{0}} \label{de}%
\end{equation}
For $\varepsilon\in\lbrack0$, $\overline{\varepsilon}[$, we denote by
$R(\epsilon)>0$ its smaller root. It is easy to see that R($\varepsilon)$ is
an increasing function of $\varepsilon.$

\paragraph{Estimates}

\begin{prop}
\label{propconv} On a compact Riemannian manifold (V,g), under the assumptions
1-2 above, for $\varepsilon>0$, the sequence\{ $u_{k}|$ k$\in\mathbb{Z}_{+}\}$
defined by the first term $u_{0}=0$ and the recurrence relation:
$\epsilon\Delta u_{k+1}+<b_{%
\operatorname{u}%
_{k}},gradu_{k+1}>_{g}+c_{u_{k}}u_{k+1}=f,$converges in the topology to the
solution u$_{\varepsilon}$ of the equation:
\[
\epsilon\Delta_{g}u_{\epsilon}+<b_{u_{\varepsilon}},gradu_{\epsilon}%
>_{g}+c_{u_{\varepsilon}}u_{\epsilon}=f
\]

Moreover for any $\varepsilon\in]0,\overline{\varepsilon}[,$
\begin{align*}
{||du||}_{\infty}\leq R(\varepsilon).
\end{align*}
For any $\varepsilon_{0}$ in $]0,\overline{\varepsilon}[$,
$\vert$%
$\vert$%
du$_{\varepsilon}||_{\infty}\leq R(\varepsilon_{0})$ for all $\varepsilon$ in
$]0,\varepsilon_{0}[.$
\end{prop}

{\large Proof:} The sequence $u_{k}$ is well defined, using the results of the
previous paragraph on the linear case. By the maximum principle:
\[
\frac{\underset{V}{\min f}}{\underset{\mathbb{R}xV}{\max c}}\leq u_{\epsilon
}\leq\frac{\underset{V}{\max f}}{\underset{\mathbb{R}\text{x}V}{\min c}}%
\]
We will prove now that $||du_{k}||_{g}$ is bounded. To do this, let us apply
the Weitzenb\"{o}ck formula to the sequence $u_{k}$,
\begin{equation}
\frac{\Delta_{g}({||du_{k}||}_{g}^{2})}{2}+||\nabla du_{k}||_{g}%
^{2}+Ricc(du_{k},du_{k})=<du_{k},d\Delta_{g}u_{k}>_{g} \label{weis}%
\end{equation}
Then
\begin{align}
&  \frac{\Delta_{g}({||du_{k}||}_{g}^{2})}{2}+||\nabla du_{k}||_{g}%
^{2}+Ricc(du_{k},du_{k})+c_{k-1}||du_{k}||_{g}^{2}+\frac{1}{2}\theta
(b_{k-1})(||du_{k}||_{g}^{2})+\nonumber\\
\lbrack &  <gradu_{k},\nabla_{gradu_{k}}b>_{g}+du_{k}\left(  \frac{\partial
b}{\partial\lambda}\right)  <du_{k},du_{k-1}>_{g}]|_{\lambda=u_{k-1}%
}=\nonumber\\
&  <df,du_{k}>_{g}-u_{k}[<dc,du_{k}>_{g}+\frac{\partial c}{d\lambda}%
<du_{k},du_{k-1}>_{g}]|_{\lambda=u_{k-1}} \label{princ}%
\end{align}
\bigskip\noindent where c$_{k}$(x)=c(u$_{k}$(x),x) and b$_{k}$(x)=b(u$_{k}%
$(x),x). In order to prove that the gradient of the sequence is bounded, we
evaluate the formula(\ref{princ}) at a maximum point P of ${||\nabla u_{k}%
||}_{g}^{2}$. Because $\Delta_{g}({||u_{k}||}_{g}^{2})(P)\geq0$ \ and
$\theta(b_{k})(||{du_{k}||}_{g}^{2}{)(P)}=0$, we get:
\begin{equation}
(c_{0}-\epsilon r_{0}-b_{0}-\beta||du_{k-1}||_{\infty})|{|du_{k}||}_{\infty
}^{2}\leq||df||_{\infty}+\frac{||f||_{\infty}}{c_{0}}[||dc||_{\infty
}+\underset{V}{\max}|\frac{\partial c}{\partial\lambda}|.||du_{k-1}||_{\infty
}] \label{appp}%
\end{equation}
Inequality (\ref{appp}) implies the following estimates for $\epsilon$ small
enough, for all $k\geq1$,
\begin{equation}
(c_{0}-b_{0}-r_{0}\epsilon-\beta{||du_{k-1}||}_{\infty})||{du_{k}||}_{\infty
}\leq f_{0}+\gamma||du_{k-1}||_{\infty} \label{majo}%
\end{equation}
Elementary properties of homographic recurrent sequences show that
${||du_{k-1}||}_{\infty}\leq R(\epsilon)$ \ implies ${||du_{k}||}_{\infty}\leq
R(\epsilon)$ for any $\varepsilon\in]0,\overline{\varepsilon}[$ (see \ref{de}
for the definition). Let us prove the result by induction: since $u_{0}=0$, we
obtain from equation (\ref{majo}) that $\max_{V_{n}}||du_{1}||_{g}\leq
\frac{f_{0}}{c_{0}-b_{0}-\varepsilon r_{0}}$. A simple computation proves that
$\frac{f_{0}}{c_{0}-b_{0}-\varepsilon r_{0}}<R(\epsilon)$ and the induction
property follows.

In order to prove that the sequence $u_{k}$ converges uniformly, we will prove
the uniform convergence of the series $w_{k}=u_{k+1}-u_{k}$. Let us prove that
each term of the series\{$w_{k}\}$ is bounded in absolute value by the terms
of a converging series . Recall that b$_{u}(x)=b(u(x),x),c_{u}(x)=c(u(x),x)$
for x$\in$V. Then from
\begin{align*}
\epsilon\Delta_{g}u_{n}+\theta(b_{u_{n-1}})u_{n}+c_{u_{n-1}}u_{n}  &  =f\\
\epsilon\Delta_{g}u_{n+1}+\theta(b_{u_{n}})u_{n+1}+c_{u_{n}}u_{n+1}  &  =f,
\end{align*}
we get:
\[
\epsilon\Delta_{g}w_{n}+\theta(b_{u_{n-1}})w_{n}+c_{u_{n-1}}w_{n}%
=\theta(b_{u_{n-1}}-b_{u_{n}})u_{n}-(c_{u_{n-1}}-c_{u_{n}})u_{n}%
\]
Let:
\[
\text{g}_{n}=\text{{}}\theta(b_{u_{n-1}}-b_{u_{n}})u_{n}-(c_{u_{n-1}}%
-c_{u_{n}})u_{n}%
\]
$\bigskip w_{n}^{2}$ satisfies:
\[
\frac{\varepsilon}{2}\Delta_{g}w_{n}^{2}+\theta(b_{u_{n}})w_{n}^{2}%
/2+c_{u_{n}}w_{n}^{2}=w_{n}g_{n}-{||dw_{n}||}_{g}^{2}%
\]
At a maximum point $P$ of $w_{n}^{2}$, $w_{n}^{2}(P)=\max_{V_{n}}w_{n}^{2}$,
$\Delta w_{n}^{2}\geq0$ and $\theta(b_{u_{n}})w_{n}^{2}=0$. Hence: $c_{u_{n}%
}(P)|w_{n}(P)|\leq g_{n}(P)$ and
\[
g_{n}(P)\leq\left(  \max_{\mathbb{R}\times V}|\frac{\partial b}{\partial
\lambda}|+\max_{\mathbb{R}\times V}|\frac{\partial c}{\partial\lambda
}|\right)  \text{ }|w_{n-1}(P)|\text{
$\vert$%
$\vert$%
}du_{n}(P)||_{g}%
\]
Since $c(\lambda,x)\geq c_{0}>0$. We obtain the estimate for $w_{n}$:
\[
\max_{V}|w_{n}|\leq\frac{1}{c_{0}}\{(\max_{\mathbb{R}\times V}|\frac{\partial
b}{\partial\lambda}|)R(\epsilon)+(\max_{\mathbb{R}\times V}|\frac{\partial
c}{\partial\lambda}|)\left(  \frac{\max|f|}{c_{0}}\right)  \}\max_{V}%
|w_{n-1}|
\]
Then:
\[
\max_{V}|w_{n}|\leq\frac{\beta R(\epsilon)+\gamma}{c_{0}}\max_{V}|w_{n-1}|
\]
Using the definition of $R(\epsilon)$ :
\begin{equation}
\beta R(\epsilon)+\gamma=c_{0}-b_{0}-\epsilon r_{0}-\frac{f_{0}}%
{R(\varepsilon)} \label{rac}%
\end{equation}
and we obtain for $\epsilon$ small enough:
\[
0<\frac{\beta R(\epsilon)+\gamma}{c_{0}}\leq\frac{c_{0}-b_{0}-\epsilon r_{0}%
}{c_{0}}-\frac{f_{0}}{c_{0}R(\epsilon)}<1
\]
This proves that the series $w_{n}$ converges geometrically in $C^{0}$
topology. Finally the sequence $u_{k}$ converges in the same topology and by
the elliptic estimate, it will converge in $C^{2,\alpha}$ topology (see
\cite{GT}) and $\ R(\epsilon)$ is a Lipschitz bound for the limit:
$\underset{V}{\max}||du_{\epsilon}||_{g}\leq R(\epsilon)$.$u_{\epsilon}$ is a
smooth solution of the equation:
\begin{equation}
\epsilon\Delta_{g}u_{\epsilon}+<b_{u_{\epsilon}},gradu_{\epsilon
}>+c_{u_{\varepsilon}}u_{\epsilon}=f \label{sinpp}%
\end{equation}

\paragraph{Weitzenb\"{o}ck Formulas}

In order to obtain regularity results,we need an estimate $\nabla
du_{\varepsilon}$ uniform in $\varepsilon$.We shall establish a generalization
of Weitzenb\"{o}ck identity which we proceed to prove. Define an extension of
the Laplacian $\Delta_{g}$ to covariant tensor fields $\omega$ as follows:
\begin{align*}
\Delta_{g}\omega_{k_{1},..k_{n}}  &  =-\nabla^{k}\nabla_{k}\omega
_{k_{1},..k_{n}}\\
\Delta_{g}\omega_{k_{1},..k_{n}}  &  =g^{ij}\left[  \left(  \nabla_{i}%
\nabla_{j}\omega\right)  _{k_{1},...,k_{n}}-\Gamma_{ij}^{k}\left(  \nabla
_{k}\omega\right)  _{k_{1},...,k_{n}}\right]
\end{align*}
For any C$^{4}-$ function u:
\begin{equation}
\Delta_{g}\nabla du=\nabla d\Delta_{g}u+\mathcal{R}_{0}(\nabla du)+\mathcal{R}%
_{1}(du) \label{wei}%
\end{equation}
where $\mathcal{R}_{0}$:$\otimes^{2}T^{\ast}V$-----%
$>$%
$\otimes^{2}T^{\star}V$ is an endomorphism of the bundle of covariant
2-tensors into itself,defined as follows:
\[
\mathcal{R}_{0}(\omega)_{i,j}=R_{i}^{l}\omega_{jl}+R_{j}^{l}\omega
_{il}+2g^{kl}R_{kij}^{\bullet\bullet\bullet m}\omega_{ml}%
\]
and $\mathcal{R}_{1}$:T*V---%
$>$
$\otimes^{2}T^{\star}V$, the vector bundle homomorphism:
\[
\mathcal{R}_{1}(\omega)_{i,j}=\nabla_{i}R_{j}^{l}\omega_{l}+\nabla_{j}%
R_{i}^{l}\omega_{l}+g^{kl}\nabla_{k}R_{ij}\omega_{l}%
\]
R$_{ij}$ is the Ricci curvature R$_{ni\cdot j}^{\cdot\cdot n\cdot}$ ,
R$_{j}^{i}=g^{in}R_{nj}=R_{jn\cdot\cdot}^{\cdot\cdot ni}$.

\noindent Then the following two facts are easy to check:(i) $\mathcal{R}_{0}
$ maps symmetric tensors into themselves, (ii) the image of $\ \mathcal{R}%
_{1}$is made up of symmetric tensors. Multiplying equation(\ref{wei}) scalarly
by$\nabla du,$a simple computation shows that :
\begin{equation}
\frac{1}{2}\Delta_{g}||\nabla du||_{g}^{2}+||\nabla^{2}du||_{g}^{2}=<\nabla
d\Delta_{g}u,\nabla du>_{g}+<\mathcal{R}_{0}(\nabla du)+\mathcal{R}%
_{1}(du),\nabla du>_{g}. \label{weisc}%
\end{equation}
The relation (\ref{weisc})is a generalization of the classical
relation(\ref{weis}).

\subsection{Regularity estimates}

\begin{prop}
\bigskip Using the notations of Proposition(\ref{propconv}), under assumptions
1-2,there exists a $\widehat{\varepsilon}>0$ such that the function of
$\epsilon,||\nabla du_{\epsilon}||_{\infty},$ is bounded on any closed
subinterval of [0,$\widehat{\varepsilon}[.$
\end{prop}

\bigskip\noindent{\large Proof: }We take the exterior derivate of
equation(\ref{sinpp}),multiply the result scalarly by $\nabla du_{\varepsilon
}.$We get: and to the solution u$_{\varepsilon}$ of equation(\ref{sinpp}) and
use this equation $:$%
\[
\varepsilon\ <\nabla d\Delta_{g}u_{\varepsilon},\nabla du_{\varepsilon}%
>_{g}+<\nabla d[(\theta(b_{u_{\epsilon}})u_{\epsilon}],\nabla du_{\varepsilon
}>_{g}+(\nabla d(c_{u_{\varepsilon}}u_{\epsilon}),\nabla du_{\varepsilon}%
>_{g}=<\nabla df,\nabla du_{\varepsilon}>_{g}%
\]
To compute $\ <\nabla d\Delta_{g}u_{\varepsilon},\nabla du_{\varepsilon}>_{g}%
$,we apply the relation (\ref{weisc}):
\begin{equation}
\frac{\varepsilon}{2}\Delta_{g}||\nabla du||_{g}^{2}+\varepsilon||\nabla
^{2}du||_{g}^{2}+<\nabla d[(\theta(b_{u_{\epsilon}})u_{\epsilon}],\nabla
du_{\varepsilon}>_{g}+(\nabla d(c_{u_{\varepsilon}}u_{\epsilon}),\nabla
du_{\varepsilon}>_{g}=<\nabla df,\nabla du_{\varepsilon}>_{g}+ \label{weieq}%
\end{equation}

\[
+<\mathcal{R}_{0}(\nabla du)+\mathcal{R}_{1}(du),\nabla du>_{g}.
\]

Let us estimate the third term on the left hand side of equation(\ref{weieq}%
).We have:
\begin{equation}
\nabla d[(\theta(b_{u_{\epsilon}})u_{\epsilon}]=\theta(b_{u_{\varepsilon}%
})\nabla du_{\varepsilon}+\mathcal{B}_{2}(b_{u_{\varepsilon}};du_{\varepsilon
})+\mathcal{R}(b_{u_{\varepsilon}},du_{\varepsilon}) \label{lie}%
\end{equation}
where, for any vector field F on V,$\omega\in$T*V---%
$>$%
$\mathcal{B}_{2}($F$;\omega)\in\otimes^{2}$T*V is the vector bundle
homo\bigskip morphism defined by:
\[
\mathcal{B}_{2}\mathcal{(}\text{F;}\omega)=\omega(\nabla^{2}F(q))\text{, if
}\omega\in\text{T}_{q}^{\ast}\text{V \ \ \ \ }%
\]
and $\mathcal{R}$: TV$\times_{V}$T*V---%
$>$%
$\otimes^{2}$T*V is the homomorphism:
\[
\mathcal{R}\text{ (X,}\omega)_{ij}=-R_{ikj}^{\cdot\text{ }\cdot\text{ }%
\cdot\text{ }l}X^{k}\omega_{l}.
\]
Note that the images of both homomorphisms are contained in the space of
symmetric 2-tensors. The proof of relation(\ref{lie}) is given in Appendix 2.
Multiplying both sides of equation(\ref{lie})\ scalarly by $\nabla
du_{\varepsilon}$:
\[
<\nabla d[(\theta(b_{u_{\epsilon}})u_{\epsilon}],\nabla du_{\varepsilon}%
>_{g}=<\theta(b_{u_{\varepsilon}})\nabla du_{\varepsilon},\nabla
du_{\varepsilon}>_{g}+<\mathcal{B}_{2}\mathcal{(}b_{u_{\varepsilon}%
};du_{\varepsilon})+\mathcal{R}(b_{u_{\varepsilon}},du_{\varepsilon}),\nabla
du_{\varepsilon}>_{g}%
\]

\[
\theta(b_{u_{\varepsilon}})\nabla du_{\varepsilon}=\nabla_{b_{u_{\varepsilon}%
}}\nabla du_{\varepsilon}+\mathcal{B}_{1}(b_{u_{\varepsilon}};\nabla
du_{\varepsilon})
\]
where for any vector field F on V,$\ \mathcal{B}_{1}$ is the homomorphism:
$\omega\in\otimes^{2}$T*V---%
$>$%
$\mathcal{B}_{1}(F;\omega)\in\otimes^{2}$T*V defined as follows:
\[
\mathcal{B}_{1}(F;\omega)_{ij}=\omega_{ki}\nabla_{j}F^{k}+\omega_{kj}%
\nabla_{i}F^{k}.
\]
We have:
\[
<\theta(b_{u_{\varepsilon}})\nabla du_{\varepsilon},\nabla du_{\varepsilon
}>_{g}=<\nabla_{b_{u_{\varepsilon}}}\nabla du_{\varepsilon},\nabla
du_{\varepsilon}>_{g}+<\mathcal{B}_{1}(b_{u_{\varepsilon}};\nabla
du_{\varepsilon}),\nabla du_{\varepsilon}>_{g}%
\]%
\[
<\theta(b_{u_{\varepsilon}})\nabla du_{\varepsilon},\nabla du_{\varepsilon
}>_{g}=\frac{1}{2}\theta(b_{u_{\varepsilon})}||\nabla du_{\varepsilon}%
||_{g}^{2}+<\mathcal{B}_{1}(b_{u_{\varepsilon}};\nabla du_{\varepsilon
}),\nabla du_{\varepsilon}>_{g}%
\]%
\begin{align*}
&  <\nabla d[(\theta(b_{u_{\epsilon}})u_{\epsilon}],\nabla du_{\varepsilon
}>_{g}=\frac{1}{2}\theta(b_{u_{\varepsilon})}||\nabla du_{\varepsilon}%
||_{g}^{2}+\\
&  <\mathcal{B}_{1}(b_{u_{\varepsilon}};\nabla du_{\varepsilon})+\mathcal{B}%
_{2}\mathcal{(}b_{u_{\varepsilon}};du_{\varepsilon})+\mathcal{R}%
(b_{u_{\varepsilon}},du_{\varepsilon}),\nabla du_{\varepsilon}>_{g}%
\end{align*}
Finally we get the identity:
\[
\frac{\varepsilon}{2}\Delta_{g}||\nabla du||_{g}^{2}+\varepsilon||\nabla
^{2}du||_{g}^{2}+\frac{1}{2}\theta(b_{u_{\varepsilon})}||\nabla
du_{\varepsilon}||_{g}^{2}+c_{u_{\varepsilon}}||\nabla du_{\varepsilon}%
||_{g}^{2}+S_{1}=S_{2}+S_{3}%
\]
where:
\begin{align*}
S_{1}  &  =<\mathcal{B}_{2}\mathcal{(}b_{u_{\varepsilon}};du_{\varepsilon
})+<\mathcal{B}_{1}(\nabla du_{\varepsilon}),\nabla du_{\varepsilon}>_{g}\\
S_{2}  &  =<-\mathcal{R}(b_{u_{\varepsilon}},du_{\varepsilon})+\varepsilon
\lbrack\mathcal{R}_{0}(\nabla du_{\epsilon})+\mathcal{R}_{1}(du_{\varepsilon
})],\nabla du_{\varepsilon}>_{g}\\
S_{3}  &  =<u_{\varepsilon}\nabla dc_{u_{\varepsilon}}+dc_{u_{\varepsilon}%
}\otimes du_{\varepsilon}+du_{\varepsilon}\otimes dc_{u_{\varepsilon}},\nabla
du_{\varepsilon}>_{g}+<\nabla df,\nabla du_{\varepsilon}>_{g}%
\end{align*}

To prove that
$\vert$%
$\vert$%
$\nabla du_{\varepsilon}||_{\infty}$ is bounded we follow the same method as
the one used to prove the boundedness of
$\vert$%
$\vert$%
du$_{\varepsilon}||_{\infty}.$ Let P$\in$V be a point where the function:x$\in
V$--%
$>$%
$||\nabla du_{\varepsilon}||_{g}^{2}(x)$ attains its maximum. Then
$\frac{\varepsilon}{2}\Delta_{g}||\nabla du||_{g}^{2}(P)\geq0$ and $\frac
{1}{2}\theta(b_{u_{\varepsilon})}||\nabla du_{\varepsilon}||_{g}^{2}(P)=0$.

Hence:
\begin{equation}
c_{u_{\varepsilon}}(P)||\nabla du_{\varepsilon}||_{g}^{2}(P)\leq\sum_{n=1}%
^{3}|S_{n}(P)| \label{funin}%
\end{equation}
To estimate S$_{1},$ it is convenient to choose an orthonormal coframe field
$\omega^{1},...,\omega^{m}$\ in a neighborhood of P. Denote by e$_{1}%
$,...,e$_{m}$ the corresponding frame field. Then:
\[
\nabla du_{\varepsilon}=U_{\alpha\beta}\omega^{\alpha}\otimes\omega^{\beta}%
\]
where the matrix of functions U$_{\alpha\beta}$ is symmetric in $\alpha,\beta
$. Moreover, because U$_{\alpha\beta}$ is symmetric we can choose the coframe
field $\omega^{1},...,\omega^{m}$ so that at P:
\[
U_{\alpha\beta}(P)=0\text{ if }\alpha\neq\beta
\]%
\[
\nabla b=B_{\beta}^{\alpha}e_{\alpha}\otimes\omega^{\beta}%
\]%
\[
\mathcal{B}_{1}(\nabla du_{\varepsilon})_{\alpha\beta}=U_{\alpha\gamma
}B_{\beta}^{\gamma}+U_{\beta\gamma}B_{\alpha}^{\gamma}%
\]%
\[
<\mathcal{B}_{1}(\nabla du_{\varepsilon}),\nabla du_{\varepsilon}>_{g}%
=\sum_{\alpha,\beta,\gamma=1}^{m}U_{\alpha\beta}(U_{\alpha\gamma}B_{\beta
}^{\gamma}+U_{\beta\gamma}B_{\alpha}^{\gamma})
\]%
\[
<\mathcal{B}_{1}(\nabla du_{\varepsilon}),\nabla du_{\varepsilon}>_{g}%
(P)=\sum_{\beta,\gamma=1}^{m}U_{\beta\beta}(P)U_{\gamma\gamma}(P)(B_{\beta
}^{\gamma}+B_{\gamma}^{\beta})(P)
\]%
\[
|<\mathcal{B}_{1}(\nabla du_{\varepsilon}),\nabla du_{\varepsilon}%
>_{g}(P)|\leq b_{0}\sum_{\alpha=1}^{m}U_{\alpha\alpha}(P)^{2}=b_{0}||\nabla
du_{\varepsilon}||_{g}^{2}(P).
\]
To estimate the term $\mathcal{B}_{2}\mathcal{(}b_{u_{\varepsilon}%
},du_{\varepsilon})$, for $\omega\in T_{q}^{\ast}V$,X,Y$\in$T$_{q}$V,q$\in
V$:
\begin{align*}
\mathcal{B}_{2}\mathcal{(}b_{u_{\varepsilon}},du_{\varepsilon})[X,Y]  &
=du_{\varepsilon}((\nabla^{2}b)_{u_{\varepsilon}}[X,Y])+du_{\varepsilon
}\left(  \left(  \nabla_{X}\frac{\partial b}{\partial\lambda}\right)
_{u_{\varepsilon}}\right)  .du_{\varepsilon}(Y)+\\
&  du_{\varepsilon}\left(  \left(  \nabla_{Y}\frac{\partial b}{\partial
\lambda}\right)  _{u_{\varepsilon}}\right)  .du_{\varepsilon}%
(X)+du_{\varepsilon}\left(  \frac{\partial b}{\partial\lambda}(u_{\varepsilon
}(q),q)\right)  .\nabla du_{\varepsilon}[X,Y]+\\
&  du_{\varepsilon}\left(  \frac{\partial^{2}b}{\partial\lambda^{2}%
}(u_{\varepsilon}(q),q)\right)  .du_{\varepsilon}(X).du_{\varepsilon}(Y).
\end{align*}
Hence:
\begin{align*}
||\mathcal{B}_{2}\mathcal{(}b_{u_{\varepsilon}},du_{\varepsilon})||_{\infty}
&  =||du_{\varepsilon}||_{\infty}.||\nabla^{2}b||_{\infty}+2||du_{\varepsilon
}||_{\infty}^{2}.\left\Vert \nabla\frac{\partial b}{\partial\lambda
}\right\Vert _{\infty}+||du_{\varepsilon}||_{\infty}^{3}.\left\Vert
\frac{\partial^{2}b}{\partial\lambda^{2}}\right\Vert _{\infty}+\\
&  \beta||du_{\varepsilon}||_{\infty}.||\nabla du_{\varepsilon}||_{g}(P)
\end{align*}
and
\begin{align*}
|S_{1}(P)|  &  \leq b_{0}||\nabla du_{\varepsilon}||_{g}^{2}(P)+R(\varepsilon
)||\nabla du(P)||_{g}\{||\nabla^{2}b||_{\infty}+2R(\varepsilon)\left\Vert
\nabla\frac{\partial b}{\partial\lambda}\right\Vert _{\infty}+R(\varepsilon
)^{2}\left\Vert \frac{\partial^{2}b}{\partial\lambda^{2}}\right\Vert _{\infty
}\\
&  +\beta||\nabla du_{\varepsilon}||_{g}(P)\}
\end{align*}%
\begin{equation}
||S_{1}||_{\infty}\leq(b_{0}+\beta R(\varepsilon))||\nabla du_{\varepsilon
}||_{\infty}^{2}+BR(\varepsilon)[1+R(\varepsilon)]^{2}||\nabla du_{\varepsilon
}||_{\infty} \label{E1}%
\end{equation}
where:
\[
B=\max\{||\nabla^{2}b||_{\infty},\left\Vert \nabla\frac{\partial b}%
{\partial\lambda}\right\Vert _{\infty},\left\Vert \frac{\partial^{2}%
b}{\partial\lambda^{2}}\right\Vert _{\infty}\}
\]
Estimate of S$_{2}$(P):
\[
|S_{2}(P)|\leq\lbrack K_{1}R(\varepsilon)||b||_{\infty}+\varepsilon
(K_{2}R(\varepsilon)+K_{3}\ ||\nabla du_{\varepsilon}(P)||_{g})]||\nabla
du_{\varepsilon}(P)||_{g},
\]
where the constants K$_{1}$ and K$_{3}$ depend only on the curvature of g and
K$_{2}$ on the covariant derivatives of the Ricci curvature tensor.
\begin{equation}
||S_{2}||_{\infty}\leq\lbrack K_{1}R(\varepsilon)||b||_{\infty}+\varepsilon
(K_{2}R(\varepsilon)+K_{3}\ ||\nabla du_{\varepsilon}||_{\infty})]||\nabla
du_{\varepsilon}||_{\infty} \label{E2}%
\end{equation}
\bigskip Estimate of S$_{3}$(P):
\[
|S_{3}(P)|\leq||\nabla du||_{\infty}\{||\nabla df||_{\infty}+C(1+R(\varepsilon
))^{2}+|u(P)\frac{\partial c}{\partial\lambda}(u(P),P)|.||\nabla du||_{\infty
}\}
\]
where C is a constant depending on f and the derivatives of c up to the second
order.
\begin{equation}
||S_{3}|_{\infty}\leq||\nabla du||_{\infty}\{||\nabla df||_{\infty
}+C(1+R(\varepsilon))^{2}+\gamma||\nabla du||_{\infty}\} \label{E3}%
\end{equation}
The inequalities (\ref{funin} \ref{E1} \ref{E2} \ref{E3}) imply the estimate:
\begin{align}
c_{0}\text{
$\vert$%
$\vert$%
}\nabla du_{\varepsilon}||_{\infty}  &  \leq(b_{0}+\beta R(\varepsilon
)+\gamma+\varepsilon K_{3})\text{
$\vert$%
$\vert$%
}\nabla du_{\varepsilon}||_{\infty}+[BR(\varepsilon)+C](1+R(\varepsilon
))^{2}+\nonumber\\
&  K_{1}R(\varepsilon)||b||_{\infty}+\varepsilon K_{2}R(\varepsilon)+||\nabla
df||_{\infty} \label{ung}%
\end{align}
The relation(\ref{rac}) implies that:
\begin{align*}
\left(  \frac{f_{0}}{R(\varepsilon)}-\varepsilon\lbrack r_{0}+K_{3}]\right)
\text{%
$\vert$%
$\vert$%
}\nabla du_{\varepsilon}||_{\infty}  &  \leq[BR(\varepsilon
)+C](1+R(\varepsilon))^{2}+[K_{1}||b||_{\infty}+\varepsilon K_{2}%
]R(\varepsilon)+\\
&  ||\nabla df||_{\infty}%
\end{align*}
Hence
$\vert$%
$\vert$%
$\nabla du_{\varepsilon}||_{\infty}$ is bounded on any proper subinterval of
[0,$\widehat{\varepsilon}[$ where $\widehat{\varepsilon}$ is either
$\overline{\varepsilon}$ or the root of : $\varepsilon R(\varepsilon)$%
=$\frac{f_{0}}{r_{0}+K_{3}}$ if it exists.

\section{Study of limit when the viscosity parameter converges to zero}

We will prove here that the first order partial differential equation
(\ref{eedp}) has a Lipschitz continuous solution obtained as a limit of the
sequence $u_{\epsilon}$ when $\epsilon$ converges to zero.

\begin{theorem}
\label{thepdf}

\begin{enumerate}
\item Under the assumptions 1-2:(i) Any sequence \{u$_{\varepsilon_{n}}$%
$\vert$
$\varepsilon_{n}$--%
$>$%
0 as n$\in\mathbb{N}$, tends to $\infty\}$ contains a sub-sequence that
converges to a solution of equation (\ref{eedp}) \ of class C$^{1}$in the
C$^{1}-$topology. Hence there exists C$^{1}$ solutions to the first order
partial differential equation (\ref{eedp}). The solutions of the equation
(\ref{eedp})which are limits of sequences \{u$_{\varepsilon_{n}}$%
$\vert$
$\varepsilon_{n}$--%
$>$%
0 as n$\in\mathbb{N}$, tends to $\infty$\} are called viscosity solutions.

\item Any viscosity solution has Lipschitz continuous derivatives with a
Lipschitz bound independent of the solution.

\item For any M
$>$%
0, there exists a neighborhood $\mathcal{N(}M)$ of 0 in C $^{0,1}$(V) such
that for any f$\in\mathcal{N(}M)$,the equation(\ref{eedp}) has at most one
Lipschitz continuous solution,having M as Lipschitz bound.
\end{enumerate}
\end{theorem}

\textbf{\noindent Proof} : (i)To study the limit of $u_{\epsilon}$ when
$\epsilon$ goes to zero, we use the same type of estimates as previously. For
all $\varepsilon>0, \underset{V}{\sup}||du_{\varepsilon}||_{g}\leq
R(\varepsilon)$. It is easy to see that R:[0,$\overline{\varepsilon}$]\ ---%
$>$%
$\mathbb{R}_{+}$ is a decreasing function. So any sequence \{u$_{\varepsilon
_{n}}|$ $\varepsilon_{n}\downarrow0$ as n $\uparrow\infty$\} contains a
subsequence converging \ in the C$^{0}$-topology to a Lipschitz continuous
function u such that $\underset{V}{\sup}||du||_{g}\leq R(0).$ Any limit u of
u$_{\varepsilon}$, as $\varepsilon$ tends to 0, satisfies the weak equation
where the integrals are taken with respect to the volume measure associated to
the metric g:
\[
\forall\phi\in H_{1}(V)\,,\int_{V}[u(\theta(b_{u})\phi)+c_{u}u\phi]=\int
_{V}f\phi
\]
and almost everywhere on V, the equation:
\[
du(x).b(u(x),x)+c(u(x),x)u(x)=f(x),\ \ x\in V
\]
Because $\nabla du_{\varepsilon}$ remains bounded as $\varepsilon$ goes to
zero, we can find a subsequence of \ \{du$_{\varepsilon_{n}}\}$, that
converges in the sup norm. Then a second subsequence \{u$_{\varepsilon_{n}}\}
$ will converge in the C$^{1}$topology to a C$^{1}$solution of equation
(\ref{eedp}).

\noindent(ii)Follows from the preceding. If u is a viscosity solution, then:
\[
Lip(u)\leq\frac{R(0)\{[BR(0)+C](1+R(0))^{2}+[K_{1}||b||_{\infty}+\varepsilon
K_{2}]R(0)+||\nabla df||_{\infty}\}}{f_{0}}%
\]
\bigskip

\noindent As in the linear case we have the formula:
\begin{align}
u(x)  &  =\int_{-\infty}^{0}K(t,x)dt,\text{ \ \ x}\in V,\text{ \ where}%
\label{integ}\\
\text{ K(t,x)}  &  =\text{f(}\varphi^{u}(t,x))\exp[-\int_{t}^{0}%
c(u(\varphi^{u}(s,x)),\varphi^{u}(s,x))ds]\nonumber
\end{align}
Here \{$\varphi_{t}^{u}|$ t$\in\mathbb{R\}}$ is the flow of the lipschitz
continuous field b$_{u}$.

Now we shall prove statement (ii). Assume that f belongs to the neighborhood
$\mathcal{F(}$M) and that equation( \ref{eedp}) has two distinct solutions
u$_{0},$u$_{1}$ both lipschitz continuous with lipschitz constant M. Let
v=u$_{1}$-u$_{0}$. Then using the relation (\ref{integ}) we get:
\begin{align}
v(x)  &  =\int_{-\infty}^{0}\{[f(\varphi^{u_{1}}(t,x)-f(\varphi^{u_{0}%
}(t,x)]e^{-\int_{t}^{0}c(u_{1}(\varphi^{u1}(s,x)),\varphi^{u_{1}}%
(s,x))ds}+\nonumber\\
&  f(\varphi^{u_{0}}(t,x))[e^{-\int_{t}^{0}c(u_{1}(\varphi^{u1}(s,x)),\varphi
^{u1}(s,x))ds}-e^{-\int_{t}^{0}c(u_{0}(\varphi^{u_{0}}(s,x)),\varphi^{u_{0}%
}(s,x)))ds}]\}dt \label{dif}%
\end{align}

where $\varphi^{u_{i}}$ denotes the flow of the Lipschitz continuous field
b$_{u_{i}}$, i=0,1. To proceed we need a careful estimate of the
$\underset{x\in V}{\max}$ d$_{g}$($\varphi^{u_{2}}(t,x),\varphi^{u_{1}%
}(t,x)).$d$_{g} $ is the distance function on V defined by the Riemannian
metric g. Let u$_{r}$=(1-r)u$_{1}+$ru$_{2}$ , r$\in\lbrack0,1]$ and denote by
\{$\varphi_{t}^{r}|t\in\mathbb{R\}}$ the flow of b$_{u_{r}}$ (instead of the
more complicated notation $\varphi_{t}^{u_{r}})$. Assume temporarily that
u$_{0}$,u$_{1}$ and hence u$_{r}$ are smooth. Denote by $\psi$:$\mathbb{R}%
\times$V$\times$[0,1]---%
$>$%
TV the mapping:$\psi$(t,x,r)=$\frac{T\varphi^{r}(t,x)}{\partial r}$. $\psi$
satisfies the variations equation:
\begin{align}
\nabla_{t}\psi(t,x,r)  &  =(\nabla_{\psi((t,x,r)}b)(u_{r}(\varphi
^{r}(t,x)),\varphi^{r}(t,x))+[v(\varphi^{r}(t,x)+du_{r}(\varphi^{r}%
(t,x)).\psi(t,x,r)]\nonumber\\
&  \left(  \frac{\partial b}{\partial\lambda}(u_{r}(\varphi^{r}(t,x)),\varphi
^{r}(t,x))\right)  \label{psi}%
\end{align}
Multiplying scalarly both sides of equation (\ref{psi}) by $\psi(t,x,r),$ we
get:
\begin{align}
&  <\nabla_{t}\psi(t,x,r),\psi(t,x,r)>_{g}=(\nabla_{\psi((t,x,r)}%
b)(u_{r}(\varphi^{r}(t,x)),\varphi^{r}(t,x)),\psi(t,x,r)>_{g}+\nonumber\\
&  <\frac{\partial b}{\partial\lambda}(u_{r}(\varphi^{r}(t,x)),\varphi
^{r}(t,x)),\psi(t,x,r)>_{g}.[v(\varphi^{r}(t,x))+du_{r}(\varphi^{r}%
(t,x)).\psi(t,x,r)] \label{psisc}%
\end{align}
{\noindent}Equation(\ref{psisc}) in turn implies the inequality:
\begin{equation}
\frac{1}{2}\frac{d}{dt}||\psi||_{g}^{2}\leq b_{0}||\psi||_{g}^{2}+\beta
\lbrack||v||_{\infty}+M||\psi||_{g}]||\psi||_{g}. \label{in}%
\end{equation}
because u admits M as Lipschitz constant. Gronwall's lemma applied to the
inequality(\ref{in}) gives\ :
\[
||\psi(t,x,r)||_{g}\leq\beta||v||_{\infty}\frac{e^{t(b_{0}+\beta M)}}%
{b_{0}+\beta M}\text{ \ for all }(t,x,r)\text{ in }\mathbb{R}_{+}\times
V\times\lbrack0,1]
\]
and $\psi$(0,x,r)=$\frac{T\varphi_{0}^{r}}{\partial r}$=0, because
$\varphi_{0}^{r}$ is the identity map of V. For negative times one uses a time
reversal and changes b into -b. Finally we get:
\[
||\psi(t,x,r)||_{g}\leq\beta||v||_{\infty}\frac{e^{|t|(b_{0}+\beta M)}}%
{b_{0}+\beta M}\text{ \ for all }(t,x,r)\text{ in }\mathbb{R}\times
V\times\lbrack0,1].
\]
Because r$\in$[0,1]---%
$>$%
$\psi(t,x,r)\in V$, is a path joining $\varphi^{u_{0}}(t,x)$ to$\ \varphi
^{u_{1}}(t,x)$:
\begin{equation}
d_{g}(\varphi^{u_{2}}(t,x),\varphi^{u_{1}}(t,x))\leq\int_{0}^{1}%
||\psi(t,x,r)||_{g}dr\leq\beta||v||_{\infty}\frac{e^{|t|(b_{0}+M\beta)}}%
{b_{0}+\beta M} \label{gron}%
\end{equation}
In the case where u$_{0}$,u$_{1}$ are just Lipschitz continuous with Lipschitz
bound M, we apply lemma(\ref{extlip}) to approximate u$_{i}$,i=0,1, by a
sequence of smooth functions \{u$_{i,n}$%
$\vert$
n$\in\mathbb{N}$\}, u$_{i,n}$ with Lipschitz bound M+$\frac{1}{n}$ and take
the limit in the inequality(\ref{gron}) corresponding to u$_{i,n}$,i=0,1.

The equation (\ref{dif}) gives us the following estimate:
\begin{align*}
||v||_{\infty}  &  \leq||v||_{\infty}\{\beta Lip(f)\int_{-\infty}^{0}%
\frac{e^{|t|(b+\beta M)-c_{0}|t|}}{b_{0}+\beta M}dt+\beta||f||_{\infty}\left(
M\underset{\mathbb{R}\times V}{\max}|\frac{\partial c}{\partial\lambda
}|+\underset{\mathbb{R}xV}{\max}||dc||_{g}\right)  .\\
&  \int_{-\infty}^{0}e^{tc_{0}}\int_{t}^{0}\text{ }\frac{e^{|s|(b_{0}+M\beta
)}}{b_{0}+\beta M}dsdt+\text{\ }||f||_{\infty}\underset{\mathbb{R}\times
V}{\max}|\frac{\partial c}{\partial\lambda}|\int_{-\infty}^{0}te^{tc_{0}}dt\}
\end{align*}
\textbf{\noindent}This implies that:
\[
||v||_{\infty}\leq||v||_{\infty}.||f||_{C^{0,1}}[\frac{\beta}{(c_{0}%
-b_{0}-M\beta)b_{0}}\left(  1+M\underset{\mathbb{R}\times V}{\max}%
|\frac{\partial c}{\partial\lambda}|+\underset{\mathbb{R}xV}{\max}%
||dc||_{g}\right)  +\frac{1}{c_{0}^{2}}\underset{\mathbb{R}\times V}{\max
}|\frac{\partial c}{\partial\lambda}|]
\]
We chose as neighborhood $\mathcal{N}(M)$ the set :
\[
\{g\in C^{0,1}(V)|\text{ }||g||_{C^{0,1}}<1/\frac{\beta}{(c_{0}-b_{0}%
-M\beta)b_{0}}\left(  1+M\underset{\mathbb{R}\times V}{\max}|\frac{\partial
c}{\partial\lambda}|+\underset{\mathbb{R}xV}{\max}||dc||_{g}\right)  +\frac
{1}{c_{0}^{2}}\underset{\mathbb{R}\times V}{\max}|\frac{\partial c}%
{\partial\lambda}|\}
\]
We conclude with some remarks. \bigskip

{\large Remark 1:} The equation(\ref{eedp}) can be interpreted as follows. Let
$\mathcal{X}$ denote the vector field on $\mathbb{R} \times V$ whose first and
second components at ($\lambda$,x)$\in\mathbb{R}\times V$ are respectively
f(x)-c($\lambda$,x)$\lambda$ and b($\lambda$,x). Then a function u:V---
$>$%
$\mathbb{R}$ is a solution of equation(\ref{eedp})if and only if the graph of
u is an invariant manifold for $\mathcal{X}$. Hence one can apply the results
of\cite{HPS,FL,Ku} to show the existence of a solution u. The advantage of our
method is that it gives precise estimates insuring the existence of a solution.

{\large Remark 2:} Our result about the existence of a solution for equation(
\ref{eedp}) states that the perturbation" $<b_{u},gradu>+c_{u}=f$ " of the
equation "$<b_{u},gradu>+c_{u}=0$" has a solution near the obvious solution 0
of "$<b_{u},gradu>+c_{u}=0$" provided that the perturbation f is small enough.
Hence our results (as well as those of the papers quoted in Remark 1) are
purely local.

{\large Remark 3:} An interesting question albeit a difficult one is to study
the structure of the set of solutions of equation(\ref{eedp}), in particular
to find the number of solutions if the set is discrete. Indeed, in the linear
case, there is a unique solution and the previous theorem \ref{thepdf} proves
that there exists a unique regular solution in a small neighborhood of the
zero solution. Outside, it is not clear how multiple solutions will appear. Is
it possible to give conditions on the recurrent sets of the vector field that
would imply a finite or countable number of solutions u.

{\large Remark 4:} It is easy to adapt our methods to the obtain solutions of
the equations:%
\begin{align}
\epsilon\Delta_{g}u_{\epsilon}(x)+  &  <b(u_{\varepsilon}%
(x),x),gradu_{\varepsilon}(x)>_{g}+c(u_{\epsilon}(x),x)=0\text{,x}\in
\text{V}\nonumber\\
&  <b(u(x),x),gradu(x)>_{g}+c(u(x),x)=0,\text{x}\in\text{V} \label{pdenl}%
\end{align}
provided that an approximate solution w:V---%
$>$%
$\mathbb{R},$ of equation(\ref{pdenl}) is known, such that the error is small
enough and $\underset{V}{\inf}$ $\frac{\partial c(w(x),x)}{\partial\lambda}$
is sufficiently big.

\section{ Some examples}

We will give some explicit examples that are usefull to understand how various
can be the solutions of first order partial differential equations.

\noindent\textbf{Example 1:} Consider real numbers K,$\alpha,\beta,$ K large
and, $\beta,\alpha$ small and a positive function $f$ on the 1-dimensional
torus S$^{1}$. Define the function c and the vector field b as follows:
\begin{align*}
&  c(u,x)=K+K\beta(1+\alpha\cos(x))u/2-\beta\alpha u/2\sin x\\
&  b(u,x)=(1+\beta(1+\alpha u\cos(x))\frac{\partial}{\partial x}%
\end{align*}
The equation :
\begin{equation}
\label{toruspde}b(u(x),x)\frac{\partial u}{\partial x}%
(x)+c(u(x),x)u(x)=f(x)\text{ , x}\in V
\end{equation}
is equivalent to
\begin{equation}
\frac{\partial U}{\partial x}+KU=f \label{linee}%
\end{equation}
where
\begin{equation}
U=u+\beta(1+\alpha cos(x))\frac{u^{2}}{2} \label{nss}%
\end{equation}
Because f $>0$, there exists a positive periodic solution of equation (
\ref{linee}). If K is big , so is $\underset{[0,2\pi]}{\min\text{ }c}$. Then
equation (\ref{nss}) has exactly two real regular solutions, if the quantity:
$\Delta(x)=1+2\beta(1+\alpha\cos(x))U$ is positive. It is the case if $\beta$
is small enough. In this case, one solution is positive, the other is
negative. But on the other hand, when $\Delta(x)$ vanishes, the solutions are
not differentiable any more. When $\Delta(x)$ changes sign, the solutions
becomes complex and in the real domain, there are pieces of intervals with no solutions.

In the context of this paper, how this example fit with proposition-1? Since
the solution $u$ is bounded a priori , the assumptions 1-2 are satisfied. In
this context, one of the two solutions is selected by the recurent process.

For the singular perturbation problem associated to equation(\ref{sinpp}),
what is the solution of the first order partial differential equation, that
will be selected at the limit, when the small parameter tends to zero:

By the maximum principle if the first term $u_{0}$ of the recurrent sequence
$u_{k}$ is positive, all the functions in the sequence are positive and so is
any limit of u$_{\varepsilon}$ as $\epsilon$ goes to zero. Hence the sequence
converge to a positive solution. Can one find an initial data, such that the
sequence $u_{k}$ converge to the other solution. How to describe the basin of
attraction of such solution associated to the map $k$---%
$>$%
$k+1$.

\textbf{Example 2:} In this second example, conditions 1-2 of theorem
(\ref{thepdf}) are satisfies and the quantities $c$ and $b$ are now bound with
the variable $u$. Consider indeed,
\begin{align*}
&  c(x,u)=K(1-\beta(1+\alpha\cos(x)))\frac{e^{-u^{2}}-1}{u}-\beta\alpha/2\sin
x\frac{e^{-u^{2}}-1}{u}\\
&  b(x,u)=(1+\beta(1+\alpha\cos(x))ue^{-u^{2}})\frac{\partial}{\partial x}.
\end{align*}
Equation \ref{toruspde} corresponding is equivalent, as in the previous
example to the equation \ref{linee} with
\begin{equation}
U=u-\beta/2(1+\alpha\cos x)\frac{e^{-u^{2}}-1}{u}. \label{inv}%
\end{equation}

It is easy to check that for K large, $c$ is big. The conditions 1-2 are
satisfied, if $\beta$ is chosen small enough, because the function
$g:u\rightarrow\frac{e^{-u^{2}}-1}{u}$ and its first derivative are bounded.
In that case there exists generically several solutions to equation (\ref{inv}).

\textbf{Remark:} This last example shows how the coefficients of the partial
differential equation can control the behavior of the sequences
u$_{\varepsilon}$ as $\varepsilon$ tends to zero. There exists other types of
examples which exhibit shock phenomena: this is given for example in dimension
1, with Burgers equation, with a forcing term (see \cite{JKM}).

As mentioned above, all solutions of first order partial differential
equations are not equivalent with respect to the singular perturbation process
and in general there are no criteria to distinguish among them. To our
knowledge, there is no theory about the stability of such solutions. We shall
not address these questions here.

\subsection{ Ergodic fields}

In this section, we analyze very briefly the case where the field b is ergodic
for the Riemannian measure,which implies that divb=0$\,$. In this case, we
obtain for the solution of the first order P.D.E. (\ref{edpline}), we prove an
averaging result without assuming that $u$ is differentiable. Remark that $u$
cannot be constant if the ratio $\frac{f}{c}$ is not constant.

The important question is what determine the value of the solution: is it
enough to give a value of the solution $u$ at an arbitrary point, and due to
the ergodicity, the field will spread the information ? Due to the recurrence
of the field, we expect that it should be the case, because along the
characteristic, which dense trajectories, the value at a specific point will
propagate on the manifold. This situation suggest also to study solutions of
equation (\ref{edpline}), when the field $b$ is a Hamiltonian system. The
phenomenon involves here should be really different from the case of
hyperbolic fields.

No limits are expected when t tends to infinity, since all trajectories are
dense in the manifold : the solution $u$ will oscillate between $\frac
{\underset{V}{\min}|f|}{\underset{V}{\max c}}$ and $\frac{\underset{V}{\max
}|f|}{\underset{V}{\min c}}$. Using the ergodicity, we can compute
$\lim_{t\rightarrow+\infty}\frac{1}{t}\int_{0}^{t}u(x(t))dt$ explicitly.

For simplicity, we suppose that $c$ is a constant function. Let P$\in$V and
let x(t) be the trajectory passing through P at time 0. Then:
\[
u(x(t))=u(P)e^{-\int_{0}^{t}c(x(s))ds}+\int_{0}^{t}f(x(\tau))e^{-\int_{\tau
}^{t}c(x(s))ds}d\tau
\]
Averaging u, we get
\[
\frac{1}{t}\int_{0}^{t}u(x(t))dt=\frac{1}{t}u(x(t_{0}))\int_{0}^{t}%
e^{-c(s-u)}du+\frac{1}{t}\int_{0}^{t}\int_{0}^{s}f(x(\tau))e^{-c(s-\tau)}d\tau
ds
\]
and changing the order of integration,
\begin{align*}
\frac{1}{t}\int_{0}^{t}\int_{0}^{s}f(x(\tau))e^{-c(s-\tau)}d\tau ds  &
=\frac{1}{t}\int_{0}^{t}f(x(\tau))\int_{\tau}^{t}e^{-c(s-\tau)}dsd\tau\\
&  =\frac{1}{t}\int_{0}^{t}\frac{f(x(\tau))}{c}(1-e^{-c(t-\tau)})du
\end{align*}
The last integral behaves like$\frac{1}{t}$ $\int_{0}^{t}\frac{f(x(\tau))}%
{c}d\tau$ and we conclude by using the ergodic properties of b:
\[
\lim_{t\rightarrow+\infty}\frac{1}{t}\int_{0}^{t}u(x(t))dt=\frac{1}%
{Vol(V)}\frac{\int_{V}f}{c}%
\]


\section{ When the zero order term vanishes}

In this paragraph, we emphasize the importance of the positivity of the lower
order term of the first order PDE studied in the previous section. In
particular, by considering a simple equation, we will prove that the results
proved in the previous section like theorem-\ref{thepdf} are no longer true.
Consider the linear partial differential operator on V:
\[
L_{\varepsilon}=\epsilon\Delta_{g}+\theta(b)
\]
and the associated equation:
\begin{equation}
\epsilon\Delta_{g}u_{\varepsilon}+\theta(b)u_{\varepsilon}=f, \label{eqred}%
\end{equation}
f is given function on V. Such an equation has a solution u$_{\varepsilon}$ if
and only if \ f is orthogonal to the kernel of the adjoint L$_{\varepsilon
}^{\ast}=\varepsilon\Delta_{g}-\theta(b)-div_{g}b$ \ of L$_{\varepsilon}$. On
the other hand if there is a solution, it is not unique since the kernel of
L$_{\epsilon}$ is not reduced to zero. All the constant functions, for
example, are in this kernel. In order to avoid these difficult questions we
shall replace the equation (\ref{eqred})by the following one:
\begin{equation}
\epsilon\Delta_{g}u_{\varepsilon}+\theta(b)u_{\varepsilon}+\varepsilon
u_{\varepsilon}=f \label{eqau}%
\end{equation}
Equation(\ref{eqau}) has a unique solution u$_{\varepsilon}$ for
\ $\varepsilon>0.$ We shall prove that there exists choices of $b$ and $f$
such that the sequence $u_{\epsilon}$ does not remain bounded as $\varepsilon$
tends to 0. As expected, the first order partial differential equation
(\ref{edpline}) has no bounded solutions $u$ for a generic f.

\begin{prop}
\label{ps} Suppose that $a$ is a critical point of the field b and
that:1)either the linear part B of the field b at a is symmetric positive with
respect to the Euclidean metric g$_{a}$ induced by g on T$_{a}$V, 2) or more
generally B=S+A, S,A$\in$End(T$_{a}$V), such that S is symmetric positive
definite with respect to g$_{a}$ and \ SA is antisymmetric with respect to
g$_{a}$. Assume also that $f(a)\neq0$. Then $u_{\epsilon}$ does not stay
bounded when $\epsilon$ goes to zero. More precisely, it diverges to infinity
at the point $a$.
\end{prop}

\noindent\textbf{Proof:}To begin with, we need some remarks in case 2). The
conditions on B imply: (i)The function: x$\in$T$_{a}$V---%
$>$%
$<$%
Sx,x%
$>$%
$_{g_{a}}$ is invariant under the action of the one parameter group e$^{tA}$
generated by A.(ii) The trace of A, tr$_{g}$A with respect to g$_{a}$ is zero.
In fact if A* denotes the adjoint of A with respect to the scalar product
$<$
,
$>$%
$_{g_{a}} $, SA+A*S=0. So SAS$^{-1}$+A*=0 and tr$_{g}$ (SAS$^{-1}$)+tr$_{g}%
$A*=0. But this gives: 2tr$_{g}$A=0.

The proof proceeds by contradiction : suppose that the sequence $u_{\epsilon}$
is bounded. Consider a normal geodesic coordinate system at point a, x$^{1}%
$,x$^{2}$,...,x$^{m}$:B---%
$>$%
$\mathbb{R}$, x$^{i}(a)$=0, 1$\leq i\leq m,$ the image of which is the ball
B$^{m}$(0,r), with center 0 and radius r in $\mathbb{R}^{m}.$ The rescaled
function:
\[
v_{\epsilon}(x)=u_{\epsilon}(\sqrt{\epsilon}x))
\]
defined on B$^{m}(0,\frac{r}{\sqrt[2]{\varepsilon}}),$ satisfies the following
equation:
\[
\Delta_{g_{\epsilon}}\text{v}_{\epsilon}(x)+\theta(\frac{b(\sqrt{\epsilon}%
x)}{\sqrt{\epsilon}})\text{v}_{\epsilon}(x)+\varepsilon\text{v}_{\epsilon
}(x)=f(\sqrt{\epsilon}x)\text{ , x}\in B^{m}(0,\frac{r}{\sqrt[2]{\varepsilon}%
})
\]
$g_{\epsilon}$ denotes the rescaled metric. The coefficients of this equation
are bounded. When $\varepsilon$ tends to 0, the rescaled metric $g_{\epsilon}$
converges to canonical metric on $\mathbb{R}^{m}$ and the field $\frac
{b(\sqrt{\epsilon}x)}{\sqrt{\epsilon}}$ converges to its linear part at 0,
i.e. the linear field $\sum_{i,j=1}^{m}B_{i,j}x^{j}\frac{\partial}{\partial
x^{i}}$ , B$_{i,j}$=$\frac{\partial b^{i}}{\partial x^{j}}(0)$, uniformly on
every compact in $\mathbb{R}^{m}$. For any sequence \{$\varepsilon_{n}|$
n$\in\mathbb{N\}}$ converging to 0, it is possible to extract a subsequence
still denoted by \{$\varepsilon_{n}|$ n$\in\mathbb{N\}}$ for simplicity, so
that \{$v_{\epsilon_{n}}\}$ converges also uniformly on every compact in
$\mathbb{R}^{m}$ to a bounded solution v of the following partial differential
equation(see \cite{GT}):
\begin{equation}
\Delta_{e}v+\sum_{i,j=1}^{m}B_{i,j}x^{j}\frac{\partial v}{\partial x^{i}%
}=f(0)\neq0\text{ on }\mathbb{R}^{m} \label{eucpde}%
\end{equation}
Here $\Delta_{e}=-\sum_{i=1}^{m}\frac{\partial^{2}}{(\partial x^{i})^{2}}.$
Let us now introduce the potential function: U:T$_{a}$V---%
$>$%
$\mathbb{R}$:
\[
U(x)=<Bx,x>_{g_{a}}=\frac{1}{2}\sum_{j=1}^{m}B_{ij}x^{i}x^{j}.
\]
Let $\psi$ = $\exp\frac{-U}{2}$. Equation \ref{eucpde} becomes after some
computations
\begin{equation}
\text{ }\Delta_{e}(v\psi)-v\Delta_{e}\psi=f(0)\psi\text{\ on }\mathbb{R}^{m}
\label{red}%
\end{equation}
This can also be written as:
\[
\text{ }\Delta_{e}(v\psi)=v\Delta_{e}\psi+f(0)\psi\text{\ on }\mathbb{R}^{m}%
\]
The right hand-side of this equation is a continuous function on
$\mathbb{R}^{m}$ tending to zero exponentially at $\infty$. Because the
function v$\psi$ is bounded on $\mathbb{R}^{m}$, its partial derivatives are
bounded on $\mathbb{R}^{m}$ (see\cite{GT}).

Finally consider the change of function w=v$\psi.$ Then equation(\ref{red})
becomes:
\[
\psi\Delta_{e}w-w\Delta_{e}\psi=f(0)\psi^{2}%
\]
To find a contradiction, we will integrate the last equation on a large ball
B(R) of center 0 and radius R in $\mathbb{R}^{m}$ with respect to Lebesgue
measure. Using the Green formula on equation the previous equation, we obtain
calling $d\sigma_{R}$ the area measure on the sphere $\partial B(R)$ and n the
unit exterior normal:
\[
\int_{\partial B(R)}(\psi\frac{\partial w}{\partial n}-w\frac{\partial\psi
}{\partial n})d\sigma_{R}=\int_{B(R)}f(0)\psi^{2}\text{ on }\mathbb{R}^{m}%
\]
Since w and dw are bounded, and $\psi$ is a Gaussian , the integral
$\int_{\partial B(R)}(\psi\frac{\partial w}{\partial n}-w\frac{\partial\psi
}{\partial n})d\sigma_{R}$ converges to zero as R tends to infinity but the
integral $\int_{\mathbb{R}^{m}}\psi^{2}$ exists and is not zero. Hence we
arrived at a contradiction if $f(0)\neq0$. So the sequence $u_{\epsilon}$ is
not bounded and no bounded solution for the first order PDE is obtained by
this method.

Let us prove now the proposition in the second case. Then v is solution of
\begin{equation}
\Delta v+\sum_{i,j=1}^{m}S_{i,j}x^{j}\frac{\partial v}{\partial x^{i}}%
+\sum_{i,j=1}^{m}A_{i,j}x^{j}\frac{\partial v}{\partial x^{i}}=f(0)\neq0\text{
on }\mathbb{R}^{m}\nonumber
\end{equation}
Consider now the potential U:T$_{a}$V---%
$>$%
$\mathbb{R}$ , associated to the endomorphism S, given by :
\[
U(x)=<Sx,x>_{g_{a}}=\sum_{i,j=1}^{m}\frac{S_{i,j}x^{i}x^{i}}{2}%
\]
and let $\psi=\exp\frac{-U}{2}$. \ Using the transformation v---%
$>$%
w=v$\psi$ as before, we see that equation transforms into
\begin{equation}
\psi\Delta w-w\Delta\psi=f(0)\psi^{2}-\psi^{2}\sum_{i,j=1}^{m}A_{i,j}%
x^{j}\frac{\partial v}{\partial x^{i}}\text{ on }\mathbb{R}^{m} \label{equrhs}%
\end{equation}
We follow the same steps as previously, the only difference is the existence
of the second term on the right hand-side of equation \ref{equrhs}. Integrate
both sides of the last equation on the solid ellipsoid E(R)=\{x%
$\vert$
U(x)%
$<$%
R$^{2}$\}, with respect to Lebesgue measure:
\[
\int_{E(R)}(\psi\Delta w-w\Delta\psi)=\int_{E(R)}f(0)\psi^{2}-\int_{E(R)}%
\psi^{2}\sum_{i,j=1}^{m}A_{i,j}x^{j}\frac{\partial v}{\partial x^{i}}%
\]
Using Green's formula to estimate the term on the left hand-side, we get:
\[
\int_{E(R)}(\psi\Delta w-w\Delta\psi)ds_{\partial E(R)}=\int_{E(R)}%
f(0)\psi^{2}-\int_{E(R)}\psi^{2}\sum_{i,j=1}^{m}A_{i,j}x^{j}\frac{\partial
v}{\partial x^{i}}%
\]
where $ds_{E(R)}$ denotes the area measure on the ellipsoid $\partial
E(R)=\{x|$ U(x)=R$^{2}\}.$ The second integral on the right hand-side of
equation is zero. Assuming this for the moment we get a contradiction in
exactly the same way as in the case when B is symmetric.

To see that the second integral on the right hand-side is zero, denote by
$\overrightarrow{A\text{ }}$the linear vector field on $\mathbb{R}^{m}$ having
$\sum_{j=1}^{m}A_{i,j}x^{j}$ as the i$^{th}$ component. The condition on the
matrix A implies that: $\theta(\overrightarrow{A)U}=0.$ This implies that E(R)
is invariant for the flow of $\overrightarrow{A}$. Hence:
\begin{align*}
0  &  =\int_{E(R)}\text{ }\theta(\overrightarrow{A)[}\psi^{2}vdx^{1}%
\wedge....\wedge dx^{m}]=\int_{E(R)}\{\text{ [(}\theta(\overrightarrow
{A})v)\psi^{2}+\text{v }\theta(\overrightarrow{A)(}\psi^{2})]dx^{1}%
\wedge....\wedge dx^{m}\\
&  +\psi^{2}v\text{ }\theta(\overrightarrow{A})[dx^{1}\wedge....\wedge
dx^{m}]\}
\end{align*}
But: $\ \theta(\overrightarrow{A)(}\psi^{2})=-\psi^{2}\theta(\overrightarrow
{A})U=0$ and $\ \theta(\overrightarrow{A})[dx^{1}\wedge....\wedge
dx^{m}]=trA[dx^{1}\wedge....\wedge dx^{m}]=0.$And $\theta(\overrightarrow
{A})v=\sum_{i,j=1}^{m}A_{i,j}x^{j}\frac{\partial v}{\partial x^{i}}$.

\noindent\textbf{Remarks:} In both cases 1) and 2) a is a hyperbolic repealer
for the flow of b. In fact the function U is a Liapunov function for the flow
of b at a in both cases.

Generically, when f does not vanishes with b, equation \ref{edpline} does not
have bound solutions. Using the characteristics solution and equation
\ref{simpl}, any solution if it exists, should blow up near 0, a zero of b
like \textquotedblleft$C\log||x||$, where $||x||$ is the distance from the
point x to 0.

\subsection{The gradient case with symmetries}

\bigskip Consider a gradient vector field $b=grad\phi$, then equation
\[
\epsilon\Delta_{g}u_{\epsilon}+\theta(b)u_{\epsilon}=f
\]
can be transformed into
\begin{align*}
div(\phi_{\epsilon}^{2}gradu_{\epsilon})  &  =f\phi_{\epsilon}^{2}\\
\phi_{\epsilon}  &  =e^{\phi/2\epsilon}%
\end{align*}
The integral condition imposes on a compact manifold that $\int_{V}%
f\phi_{\epsilon}^{2}=0$. Suppose that the manifold is a sphere, f is odd and
$\phi$ is even. The integral condition is then satisfied by the symmetries
assumptions. Assume that $\phi$ has a minimum where $f$ is not zero. We can
apply locally the results of Proposition-\ref{ps} to show that the sequence
$u_{\epsilon}$ is not bounded.

\subsection{Negative case}

We finish this section by mentioning an open problem concerning the
construction of a sequence of approximate solution of the following equation:
\[
\Delta_{g}u_{\epsilon}+<b,\nabla u_{\epsilon}>-\epsilon^{-2}u_{\epsilon
}=f\text{ on }V
\]
This problem is related to the linear wave equation on manifold when $b=0$. It
is known that there exists a subsequence of $u_{\epsilon}$ that concentrates
along a stable close geodesic (see Babich-Lazutkin, \cite{Babich}) of the
manifold. We ask the following question: when ${\varepsilon}$ tends to zero,
is it possible to construct a subsequence that concentrate along the limit
sets of the vector field, instead of the geodesics? moreover what are the
minimal assumptions on the vector field $b$ that insure the concentration of
the sequence of eigenfunctions ?

\section{Conclusion}

We have studied the scalar case. But our methods could be extended to study
systems of the form:
\begin{align*}
\varepsilon\Delta_{g}u+\sum_{n=1}^{m}B_{n}(u(x),x)\frac{\partial u}{\partial
x^{n}}+C(u(x),x)u(x)  &  =f(x),\text{x}\in\text{V}\\
\sum_{n=1}^{m}B_{n}(u(x),x)\frac{\partial u}{\partial x^{n}}+C(u(x),x)u(x)  &
=f(x),\text{x}\in\text{V}%
\end{align*}
where u,f:V---%
$>$%
$\mathbb{R}^{N}$ and B$_{n}$,C :$\mathbb{R}$xV---%
$>$%
End($\mathbb{R}^{N}$) are given matrix functions. C must satisfy positivity
conditions (see\cite{FL},\cite{FL1}).


\section{Appendix 1}

In this appendix we give a proof of lemma(\ref{extlip}). We embed \ (V,g)
isometrically into a Euclidean space $\mathbb{R}^{n}$ endowed with its
canonical metric. There exists a r*
$>$%
0, depending only on the normal curvatures of V in $\mathbb{R}^{n}$, such that
for any r$\in]0,$r*[, V has a tubular neighborhood (T$_{r}$,$\pi_{r}$,V) of V
in $\mathbb{R}^{n}$ of the following type:for any x$\in$V, $\pi_{r}$(x) is the
open ball of center x and radius r in the affine space N$_{x}$V, passing
through x and orthogonal to V at x. It is easy to see that there exist a
function c:[0,r*[---%
$>$%
[0,+$\infty$[, such that c(r) tends to 0 as r tends to 0 and max\{T$\pi_{r}%
$(v) $|$ v tangent to T$_{r}$ and
$\vert$%
$\vert$%
v%
$\vert$%
$\vert$%
$_{\mathbb{R}^{n}}$=1\}$\leq1+c(r).$ \ \ \ \ \ \ \ \ \ \ \ \ \ \ \ \ \ \ 

Call $\widehat{k_{r}}$ : T$_{r}$---%
$>$%
$\mathbb{R}$, the function k$\circ\pi_{r}.$ As a function on the manifold
T$_{r}$, $\widehat{k}_{r}$ is lipshitz continuous with lipschitz bound
M(1+c(r)). This means that for any couple x,y in T$_{r}$, for any C$^{1}$curve
$\gamma:[0,1]$---%
$>$%
T$_{r}$, such that $\gamma(0)$=x, $\gamma(1)$=y,
$\vert$%
$\vert$%
$\widehat{k}_{r}(x)-\widehat{k}_{r}(y)||_{\mathbb{R}^{m}}\leq$
M(1+c(r))$\times$ length of $\gamma.$ Rademacher's theorem implies that
$\widehat{k}_{r}$ belongs to the Sobolev space W$^{1,\infty}(T_{r})$.

Choose an r such that 2r%
$<$%
r* and that Mc(2r)%
$<$%
$\varepsilon.$ Let a:$\mathbb{R}^{n}$---%
$>$%
[0,+$\infty$[ be a C$^{\infty}$ function with support in the ball of center 0
and radius r in $\mathbb{R}^{n}$ and such that $\int_{\mathbb{R}^{n}}%
$a(x)dx=1. Consider the convolution $\widehat{h}_{r}$:T$_{r}$---%
$>$%
$\mathbb{R}$:
\[
\widehat{h}_{r}(x)=\int_{\mathbb{R}^{n}}\widehat{k}_{r}(y)a\left(  x-y\right)
dy=\int_{\mathbb{R}^{n}}\widehat{k}_{r}(x-z)a(z)dz
\]
Clearly $\widehat{h}_{r}$ is C$^{\infty}$. If x,y$\in$T$_{r}$ and
$\vert$%
$\vert$%
x-y%
$\vert$%
$\vert$%
$_{\mathbb{R}^{n}}$%
$<$%
r, the straight line segment joining x-z to y-z is contained in T$_{2r}$, for
any z such that \
$\vert$%
$\vert$%
z%
$\vert$%
$\vert$%
$_{\mathbb{R}^{n}}$%
$<$%
r. Then for all these z ,%
$\vert$%
$\vert$%
$\widehat{k}_{r}(x-z)-\widehat{k}_{r}(y-z)||_{\mathbb{R}^{n}}\leq
(M+\varepsilon)||x-y||_{\mathbb{R}^{n}}$. Hence
$\vert$%
$\vert$%
$\widehat{h}_{r}(x)-$ $\widehat{h}_{r}(y)$%
$\vert$%
$\vert$%
$_{\mathbb{R}^{n}}\leq(M+\varepsilon)||x-y||_{\mathbb{R}^{n}}.$ This implies
that
$\vert$%
$\vert$%
d$\widehat{h}_{r}||\leq M+\varepsilon$.

If x$\in T_{r}$ and z is such that \
$\vert$%
$\vert$%
z%
$\vert$%
$\vert$%
$_{\mathbb{R}^{n}}$%
$<$%
r, the segment joining x and x-z is contained in T$_{2r}$ and $\widehat
{||k}_{r}(x-z)-\widehat{k}_{r}(x)||_{\mathbb{R}^{n}}\leq(M+\varepsilon)r.$
Hence:
\[
||\widehat{h}_{r}(x)-\widehat{k}_{r}(x)||_{\mathbb{R}^{n}}\leq(M+\varepsilon
)r\text{ for all x}\in\text{T}_{r},%
\]
By taking r sufficiently small the restriction h of $\ \widehat{h_{r}}$ to V
will belong to $\mathfrak{U}$ \ and will have M+$\varepsilon$ as Lipschitz
bound.

\section{ Appendix 2}

In this second appendix, we give the computations related to formula
\ref{lie}. Let $\omega$ be a 1-form. We have:%
\[
\theta(b)\omega=\nabla_{b}\omega+A_{b}\omega
\]

where:%
\[
A_{b}\omega\lbrack X]=\omega\lbrack\nabla_{X}b]
\]

for any vector field X.Hence:%
\[
\nabla\theta(b)\omega=\nabla\nabla_{b}\omega+\nabla A_{b}\omega
\]

We compute both terms on the right hand side using a coordinate system
(x$^{1}$,...,x$^{m}).$%
\begin{align*}
\nabla\nabla_{b}\omega &  =\nabla_{i}\nabla_{b}\omega_{j}dx^{i}\otimes
dx^{j}\\
\nabla_{i}\nabla_{b}\omega &  =\nabla_{b}\nabla_{i}\omega+\nabla_{i}%
b^{j}\nabla_{j}\omega+R(e_{i},b)\omega\\
R(e_{i},b)\omega &  =-R_{ikj}^{\bullet\bullet\bullet l}b^{k}\omega_{l}dx^{j}\\
\nabla_{i}\nabla_{b}\omega_{j}  &  =\nabla_{b}\nabla_{i}\omega_{j}+\nabla
_{i}b^{k}\nabla_{k}\omega_{j}-R_{ikj}^{\bullet\bullet\bullet l}b^{k}\omega
_{l}\\
\nabla\nabla_{b}\omega &  =\nabla_{b}\nabla\omega+\nabla b^{j}\nabla_{j}%
\omega-R_{ikj}^{\bullet\bullet\bullet l}b^{k}\omega_{l}dx^{i}\otimes dx^{j}%
\end{align*}

$A_{b}\omega=\omega_{i}\nabla_{j}b^{i}dx^{j}$. So it is the contraction of the
1-covariant tensor $\omega$ and the (1,1) variant tensor $\nabla b$and:%
\[
\nabla_{i}(A_{b}\omega)_{j}=\nabla_{i}\omega_{k}\nabla_{j}b^{k}+\omega
_{k}\nabla_{i}\nabla_{j}b^{k}%
\]

\bigskip Finally:%

\[
\nabla_{i}\theta(b)\omega_{j}=\nabla_{b}\nabla_{i}\omega_{j}+\nabla_{i}%
\omega_{k}\nabla_{j}b^{k}+\nabla_{i}b^{k}\nabla_{k}\omega_{j}-R_{ikj}%
^{\bullet\bullet\bullet l}b^{k}\omega_{l}+\omega_{k}\nabla_{i}\nabla_{j}b^{k}%
\]

But we have, if $\alpha=\alpha_{ij}dx^{i}\otimes dx^{j}$ is a 2- covariant
tensor:%
\begin{align*}
(\theta(b)\alpha)_{ij}  &  =\theta(b)\alpha_{ij}+\alpha_{kj}\frac{\partial
b^{k}}{\partial x^{i}}+\alpha_{ik}\frac{\partial b^{k}}{\partial x^{j}}\\
(\theta(b)\alpha)_{ij}  &  =(\nabla_{b}\alpha)_{ij}+\alpha_{kj}\nabla_{i}%
b^{k}+\alpha_{ik}\nabla_{j}b^{k}%
\end{align*}

Applying this formula to $\alpha=\nabla\omega:$%
\[
(\theta(b)\nabla\omega)_{ij}=\nabla_{b}\nabla_{i}\omega_{j}+\nabla_{i}%
\omega_{k}\nabla_{j}b^{k}+\nabla_{k}\omega_{j}\nabla_{i}b^{k}%
\]

We get:%
\[
\nabla\theta(b)\omega=\theta(b)\nabla\omega+\mathcal{R(}b;\omega
)+\mathcal{B}(b;\omega)
\]

where:%
\begin{align*}
\mathcal{R(}b;\omega)_{ij}  &  =-R_{ikj}^{\bullet\bullet\bullet l}b^{k}%
\omega_{l}\\
\mathcal{B}(b;\omega)_{ij}  &  =\omega_{k}\nabla_{i}\nabla_{j}b^{k}%
=\omega(\nabla^{2}b)
\end{align*}

\end{document}